\newif\ifsubmode 
\submodefalse

\ifsubmode
\documentclass[10pt,preprint]{aastex}
  \received{}
  \revised{}
  \accepted{}
\else
  \documentclass[iop,numberedappendix]{emulateapj}   \usepackage{apjfonts}
\fi

\usepackage{natbib}
\usepackage{epsf}
\usepackage{color}
\usepackage{amsmath}
\usepackage{graphicx}
\newcommand{\hst}{\textit{HST}}

\newcommand{\herschel}{\textit{Herschel}}
\newcommand{\spitzer}{\textit{Spitzer}}

\newcommand{\acsb}{\hbox{$B_{435}$}}
\newcommand{\acsv}{\hbox{$V_{606}$}}
\newcommand{\acsi}{\hbox{$i_{775}$}}
\newcommand{\acsz}{\hbox{$z_{850}$}}

\newcommand{\wfcj}{\hbox{$J_{125}$}}
\newcommand{\wfch}{\hbox{$H_{160}$}}

\newcommand{\ks}{\hbox{$K_s$}}

\newcommand{\lsim}{\lesssim}
\newcommand{\gsim}{\gtrsim}

\newcommand{\eg}{e.g.}

\newcommand{\msol}{\hbox{$M_\odot$}}

\newcommand{\lsol}{\hbox{$L_\odot$}}

\newcommand{\lir}{\hbox{$L_{\mathrm{IR}}$}}

\newcommand{\jone}{\hbox{$J_1$}}
\newcommand{\jtwo}{\hbox{$J_2$}}
\newcommand{\jthree}{\hbox{$J_3$}}
\newcommand{\hs}{\hbox{$H_s$}}
\newcommand{\hl}{\hbox{$H_l$}}

\newcommand{\reff}{\hbox{$r_\mathrm{eff}$}}

\newcommand{\sersic}{S\'ersic}
\newcommand{\nsersic}{\hbox{$n_s$}}
\newcommand{\zfourge}{ZFOURGE}

\ifsubmode
\else

\fi
\newcommand{\ed}[1]{{#1}}
\newcommand{\edtoo}[1]{{#1}}

\shorttitle{EVOLUTION OF M$^\ast$ GALAXY PROGENITORS}
\shortauthors{PAPOVICH ET AL.}

\begin{document}

%\slugcomment{\it Draft Version \today, \ampmtime}
\slugcomment{\it  accepted for publication in ApJ; draft version 2014 December 8}
%\slugcomment{Accepted for Publication in the Astrophysical Journal}
%
%\title{ZFOURGE/CANDELS: ON THE EVOLUTION OF M$^\ast$ GALAXY PROGENITORS FROM Z=3 TO 0.5\altaffilmark{$\ast$}}
\ifsubmode
\title{ZFOURGE/CANDELS: ON THE EVOLUTION OF M$^\ast$ GALAXY PROGENITORS FROM Z=3 TO 0.5$^\ast$}
\else
\title{ZFOURGE/CANDELS: ON THE EVOLUTION OF M$^\ast$ GALAXY PROGENITORS FROM Z=3 TO 0.5\altaffilmark{$\ast$}}
\fi

\author{\sc C.~Papovich\altaffilmark{1,2}, 
I.~Labb\'e\altaffilmark{3}, 
R.~Quadri\altaffilmark{1,2,28},
 V.~Tilvi\altaffilmark{1,2}, 
P.~Behroozi\altaffilmark{8},
E.~F.~Bell\altaffilmark{4}, 
K.~Glazebrook\altaffilmark{5},  
L.~Spitler\altaffilmark{6,7},
  C.~M.~S.~Straatman\altaffilmark{3},  
K.-V.~Tran\altaffilmark{1,2},  
M.~Cowley\altaffilmark{6}, 
R.~Dav\'e\altaffilmark{9,10,11},
 A.~Dekel\altaffilmark{12}, 
M.~Dickinson\altaffilmark{13},
  H.~C.~Ferguson\altaffilmark{8}, 
S.~L.~Finkelstein\altaffilmark{14},
E.~Gawiser\altaffilmark{15}, 
H.~Inami\altaffilmark{13},
S.~M.~Faber\altaffilmark{16},
  G.~G.~Kacprzak\altaffilmark{5,29},  
L.~Kawinwanichakij\altaffilmark{1,2}, 
D.~Kocevski\altaffilmark{17},
  A.~Koekemoer\altaffilmark{8}, 
D.~C.~Koo\altaffilmark{16},
 P.~Kurczynski\altaffilmark{15}, 
J.~M.~Lotz\altaffilmark{8}, 
Y.~Lu\altaffilmark{18},
R.~A.~Lucas\altaffilmark{8},
  D.~McIntosh\altaffilmark{19},
N.~Mehrtens\altaffilmark{1,2}, 
B.~Mobasher\altaffilmark{20}, 
A.~Monson\altaffilmark{21}, 
G.~Morrison\altaffilmark{22,23}, 
  T.~Nanayakkara\altaffilmark{5}, 
S. E. Persson\altaffilmark{21}, 
B.~Salmon\altaffilmark{1,2}, 
R.~Simons\altaffilmark{24},
  A.~Tomczak\altaffilmark{1,2}, 
P.~van~Dokkum\altaffilmark{25}, 
B.~Weiner\altaffilmark{26}, 
S.~P.~Willner\altaffilmark{27}
}
%
%\ifsubmode \altaffiltext{1}{George P.\ and Cynthia Woods Mitchell
%Institute for Fundamental Physics and Astronomy, and  Department of
%Physics and Astronomy, Texas A\&M University, College Station, TX,
%77843-4242; papovich@physics.tamu.edu} \else

\altaffiltext{$\ast$}{This paper contains data gathered with the 6.5 meter
  Magellan Telescopes located at Las Campanas Observatory, Chile.}
\affil{$^{1}$George P.\ and Cynthia Woods Mitchell Institute for
  Fundamental Physics and Astronomy, Texas A\&M University, College Station, TX, 77843-4242 USA}
\affil{$^{2}$Department of Physics and Astronomy, Texas A\&M University, College
Station, TX, 77843-4242 USA; papovich@tamu.edu}
\affil{$^{3}$Leiden Observatory, Leiden University, P.O. Box 9513, 2300 RA Leiden, The Netherlands}
\altaffiltext{28}{Mitchell Astronomy Fellow}
\affil{$^{4}$ Department of Astronomy, University of Michigan, Ann Arbor, MI 48109, USA}
\affil{$^{5}$Centre for Astrophysics \& Supercomputing, Swinburne
  University, Hawthorn, VIC 3122, Australia}
\affil{$^{6}$Department of Physics \& Astronomy, Macquarie University, Sydney, NSW 2109, Australia}
\affil{$^{7}$Australian Astronomical Observatory, 105 Delhi Rd, Sydney, NSW 2113, Australia}
\affil{$^{8}$Space Telescope Science Institute, 3700 San Martin Dr.,
  Baltimore, MD 21218 USA}
\affil{$^{9}$University of the Western Cape, Bellville, Cape Town
  7535, South Africa}
\affil{$^{10}$South African Astronomical Observatories, Observatory,
  Cape Town 7925, South Africa}
\affil{$^{11}$African Institute for Mathematical Sciences, Muizenberg, Cape Town 7945, South Africa}
\affil{$^{12}$Center of Astrophysics and Planetary Sciences, Racah
  Institute of Physics, The Hebrew University of Jerusalem, Jerusalem
  91904, Israel}
\affil{$^{13}$National Optical Astronomy Observatory, 950 N.\ Cherry
  Ave., Tucson, AZ 85721 USA}
\affil{$^{14}$Department of Astronomy, University of Texas, Austin, TX 78712, USA}
\affil{$^{15}$Department of Physics \& Astronomy, Rutgers University,
  Piscataway,  NJ 08854 USA}
\altaffiltext{29}{Australian Research Council Super Science Fellow}
\affil{$^{16}$University of California Observatories/Lick Observatory,
University of California, Santa Cruz, CA 95064, USA}
\affil{$^{17}$Department of Physics and Astronomy, University of
  Kentucky, Lexington, KY 40506, USA}
\affil{$^{18}$Kavli Institute for Particle Astrophysics and Cosmology, Stanford University, Stanford, CA 94305, USA}
\affil{$^{19}$Department of Physics, University of Missouri-Kansas City,
5110 Rockhill Road, Kansas City, MO 64110, USA}
\affil{$^{20}$Department of Physics and Astronomy, University of
  California, Riverside, CA 92521, USA}
\affil{$^{21}$Carnegie Observatories, Pasadena, CA 91101, USA}
\affil{$^{22}$ Institute for Astronomy, University of Hawaii, 
Manoa, Hawaii 96822-1897 USA}
\affil{$^{23}$ Canada-France-Hawaii Telescope Corp.,
Kamuela, Hawaii 96743-8432, USA}
\affil{$^{24}$Department of Physics \& Astronomy, The Johns Hopkins
  University, Baltimore, MD 21218 USA}
\affil{$^{25}$Department of Astronomy, Yale University, New Haven, CT
  06520, USA}
\affil{$^{26}$Steward Observatory, University of Arizona, Tucson, AZ
  85721 USA}
\affil{$^{27}$Harvard-Smithsonian Center for Astrophysics, Cambridge,
  MA 02138 USA}
%\fi

%%%%%%%%%%%%%%%%%%%%%%%%%%%%%%%%%%%%%%%%%%%%%%%%%%%%%%%

%\setcounter{footnote}{10}

\begin{abstract}  

\noindent 
Galaxies with stellar masses near $M^\ast$ contain the majority of
stellar mass in the universe, and are therefore of special interest in
the study of galaxy evolution. The Milky Way (MW) and Andromeda (M31)
have present day stellar masses near $M^\ast$, at
$5\times~10^{10}$~\msol\ (defined here to be MW-mass) and
$10^{11}$~\msol\ (defined to be M31-mass).  We study the typical
progenitors of these galaxies using \zfourge, a deep medium-band
near-IR imaging survey, which is sensitive to the progenitors of these
galaxies out to $z\sim~3$.  We use abundance-matching techniques to
identify the main progenitors of these galaxies at higher redshifts.
We measure the evolution in the stellar mass, rest-frame colors,
morphologies, far-IR luminosities, and star-formation rates combining
our deep multiwavelength imaging with near-IR \hst\ imaging from
CANDELS, and \spitzer\ and \herschel\ far-IR imaging from GOODS-H and
CANDELS-H. The typical MW-mass and M31-mass progenitors passed through
the same evolution stages, evolving from blue, star-forming disk
galaxies at the earliest stages, to redder dust-obscured IR-luminous
galaxies in intermediate stages, and to red, more quiescent galaxies
at their latest stages.  The progenitors of the MW-mass galaxies
reached each evolutionary stage at later times (lower redshifts) and
with stellar masses that are a factor of 2--3 lower than the
progenitors of the M31-mass galaxies.   The process driving this
evolution, including the suppression of star-formation in present-day
$M^\ast$ galaxies requires an evolving stellar-mass/halo-mass ratio
and/or evolving halo-mass threshold for quiescent galaxies.  The
effective size and star-formation rates imply that the baryonic
cold--gas fractions drop as galaxies evolve from high redshift to
$z\sim 0$ and are strongly anticorrelated with an increase in the
\sersic\ index.  Therefore, the growth of galaxy bulges in $M^\ast$
galaxies corresponds to a rapid decline in the galaxy gas fractions
and/or a decrease in the star-formation efficiency. 

\end{abstract}
 
\keywords{ galaxies: evolution --- galaxies: high-redshift --- galaxies:  structure} 

%%%%%%%%%%%%%%%%%%%%%%%%%%%%%%%%%%%%%%%%%%%%%%%%%%%%%%%%%%%%%%%%%%%%%%

\section{INTRODUCTION}

\setcounter{footnote}{29}

Studying the formation of galaxies with stellar masses like the Milky Way (MW) and
Andromeda (M31) provides insight into the formation of large
galaxies and the most common locations of stars in the present Universe.
Galaxies with these masses constitute the majority of the bright galaxy
population in the local universe:  by number they represent
70\% of the intermediate-mass galaxy population (ranging from $3\times
10^{10}$ to $3\times 10^{11}$ \msol), and they contain more than
two-thirds of the present-day stellar mass density when integrated
over the entire mass function \citep[\eg,][]{hammer07}.   Despite the
fact that these galaxies are so ubiquitous and common, our knowledge
of the formation of these galaxies, such as the MW, is still largely
incomplete \citep{rix13}.

Both the MW and M31 have stellar masses very near the present-day
values of $M^\ast$, the characteristic stellar mass of the galaxy
stellar mass function, which is described by the well-known Schechter
function \ed{\citep[see \eg,][and references therein]{bell03b,baldry08,ilbe13,mous13,muzz13c,tomc14}}, 
\begin{equation}
\phi(M_\ast)\, dM_\ast = \phi^\ast \left( \frac{M_\ast}{M^\ast} \right)^\alpha
\exp(-M_\ast/M^\ast) \frac{dM_\ast}{M^\ast}. 
\end{equation} 
$M^\ast$ is a fundamental parameter and corresponds to the point where
the stellar-mass function transitions from a power-law in stellar mass
to an exponentially declining cut-off.\footnote{Although there is
evidence that the galactic stellar mass function is better represented
as a double-Schechter function these double-Schechter functions are
typically consistent with a single $M^\ast$ value, at least for
$z$$<$2 e.g., \citet{baldry08,tomc14}.}     As illustrated in
figure~\ref{fig:mf}, $M^\ast$ sits near the peak of the
stellar-mass-distribution function (the product of the stellar mass
function and the stellar mass):  $M^\ast$ is the
``mode'' of the stellar-mass-density distribution function.  Therefore
a typical star (such as the Sun) most commonly resides in galaxies of
this stellar mass at present \citep{vandokkum13}.\footnote{At any
redshift the most common location of stars will be in galaxies around
the value $M^\ast(z)$. Because $M^\ast(z)$ does not evolve strongly
with redshift \ed{\citep[see \eg,][]{ilbe13,muzz13c,tomc14}}, it is
only at  present ($z$=0) that galaxies with masses like the MW and M31
are the most common locations of stars.  As we discuss in this paper,
the progenitors of MW-mass and M31-mass galaxies are lower than
$M^\ast$ at earlier times (higher redshift), and therefore the
progenitors of the MW-mass and  M31-mass galaxies are \textit{not} the
most common locations of stars at earlier epochs. }   By studying the
evolution of present-day $M^\ast$ galaxies, we are able to learn about
the most common sites of stars in the present-day Universe\ed{, including
the formation of the MW and M31.}   

\ifsubmode
\begin{figure} 
\epsscale{0.75}
\else
\begin{figure}  
\epsscale{1.05} 
\fi 
\plotone{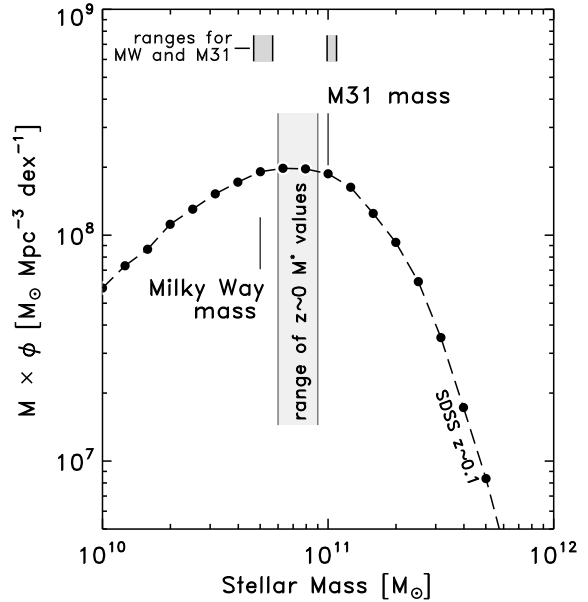}\caption{Stellar mass density distribution derived from
the product of stellar mass and the stellar mass function at $z\sim
0.1$ \citep[see also van Dokkum et al.\ 2013]{mous13}.    These
distributions peak around $M^\ast$, the characteristic mass of the
Schechter function, and the large shaded swath indicates the range of
low-redshift $M^\ast$ values in the literature.      Our adopted values for the
mass of MW-mass galaxies (stellar mass of $5\times 10^{10}$~\msol) and
M31-mass galaxies ($10^{11}$~\msol) are indicated in the figure.
These are consistent with measurements of the MW and M31 proper, where
the smaller shaded regions near top of the figure show the values for
the MW and M31 from \citet{mutch11}.   Our adopted values for
the MW-mass and M31-mass galaxies span the full range of $M^\ast$,
allowing us to study the range of galaxies with masses near the mode
of the stellar-mass density distribution, and this includes possible
formation histories of our own Galaxy.  }\label{fig:mf}
\epsscale{1}
 \ifsubmode
\end{figure}
\else
\end{figure}
\fi

The complex evolution of $M^\ast$ galaxies has been the focus of galaxy formation models within cosmological
simulations, which include the properties of the dark matter, gas
accretion, and feedback
\citep[\eg,][]{bour07b,elme08,agertz09,dekel09b,martig09,martig10}.
These studies include the effects of cold gas flows,
star-forming clump formation and migration, and violent disk
instabilities on bulge formation
\citep{ceve10,ceve12,sales12,zava12,dekel09b,dekel13,dekel14}.   These
models make predictions for the relation between stellar mass growth,
structural evolution, and the evolution of the star-formation rate
(SFR), gas accretion rate, and gas fraction for galaxies with masses
of the MW and M31.

Comparing the predictions from models of $M^\ast$ galaxy formation to
data has been hindered by observational limitations.   The models
predict that the progenitors of these galaxies should have stellar
masses of $\lsim 10^{10}$~\msol\ at $z \gsim 2$
\citep[\eg,][]{derossi09,moster13,behr13c}, and surveys typically with
the depth required to be complete to this stellar mass have very
small fields that lack the cosmic volume to trace the progenitors of
these galaxies across cosmic time in a homogenous dataset \citep[e.g.,
to be complete for galaxies to this limiting stellar mass at this
redshift requires typically $K_\mathrm{AB} \gsim 24$~mag, see for
example,][]{bass13}.   Furthermore, although simulations track the
formation of individual MW-like galaxies over long baselines in time,
this is clearly not possible in observational surveys.  Rather, to make
empirical constraints requires that we identify galaxies at high
redshift that are statistically similar to the progenitors of
nearby galaxies observed over a range of redshift.

Recent surveys, using very deep near-IR imaging have begun to study
the evolution of present-day galaxies such as the MW.  Using data from
the 3D--HST and CANDELS surveys, \citet{vandokkum13} studied the
assembly history and evolution of structural properties of galaxies
with a present-day mass of a MW-sized galaxy (assuming $M_\ast \simeq
5\times 10^{10}$~\msol) by assuming the main progenitors of these
galaxies have constant (comoving) number density at higher redshift.
They found that $\sim$90\% of the stellar mass in these galaxies has
been built since $z\sim 2.5$ without any significant merging.
\citet{patel13b} focused on star-forming progenitors of galaxies with
a present-day stellar mass of $\simeq 3\times 10^{10}$~\msol, based on
the evolution of galaxies along the star-forming ``main sequence''
\citep[\eg,][]{noeske07a,karim11,leit12}.  Both the studies of van
Dokkum et al.\ and Patel et al.\ found a peak SFR$\simeq$10--15~\msol\
yr$^{-1}$ at $z$$\sim$1--2 for these galaxies, where most of this
stellar mass growth occurred at nearly the same rate at all radii with
no evidence for inside-out growth, at least for progenitors at $z >
0.6$.  

However, it remains unclear how this evolution proceeded, and what
physical processes regulated it.   Clearly, if star-formation
dominated the formation of $M^\ast$ galaxies as suggested by
\citet{vandokkum13} and \citet{patel13b}, then their growth was heavily
dependent on the evolution of their cold gas supply and their
gas-accretion histories \citep[the SFR is expected to track
the gas accretion history, see, \eg,][]{agertz09,dekel13}.  Therefore,
understanding the evolution of the galaxies' gas is paramount.
Clearly, the processes driving galaxy formation and assembly depend on
galaxy mass \citep[\eg,][]{moster13}.  Because these
processes are complex, the assembly histories of the progenitors of
present-day $M^\ast$-mass galaxies should have a large variation that
depends on the mass of the galaxies' main progenitors
\citep[\eg,][]{behr13c}.   Therefore, to study how the formation of
$M^\ast$ galaxies proceeded, it is important to consider how the
physical properties of these galaxies evolved as a function of stellar
mass and redshift. 

Here we use data from a combined set of deep surveys to study the
evolution of progenitors of $M^\ast$ galaxies.  The combined datasets
here include data from the {\sc FourStar} Galaxy Evolution (\zfourge)
survey, the Cosmic Near-IR Deep Extragalactic Legacy Survey (CANDELS),
including \spitzer\ and \herschel\ imaging from CANDELS-Herschel
(CANDELS-H) and the Great-Observatories Origins Deep Survey-Herschel
(GOODS-H).   

The outline for this paper is as follows.  \S~2 discusses the
properties of present-day $M^\ast$ galaxies and how they relate to the
MW and M31.  \S~3 describes the \zfourge, CANDELS \hst, \spitzer, and
\herschel\ datasets, it discusses the derivation of physical
properties such as photometric redshifts, stellar masses, rest-frame
colors, sizes, and \sersic\ indices. \S~4 discusses the selection of
$M^\ast$ galaxy progenitors (including the progenitors of MW-mass and
M31-mass galaxies), incorporating the expected galaxy growth from
abundance matching methods.  \S~5 discusses the color of the $M^\ast$
galaxy progenitors, and \S~6 discusses the evolution of the galaxy
morphologies.  \S~7 describes the stacked far-IR data from the
$M^\ast$ galaxy progenitor samples, and it discusses the
evolution in galaxy IR luminosities, SFRs, and implied gas fractions.
\S~8 discusses constraints on the growth of $M^\ast$ galaxy
progenitors, and shows how the combination of these independent
datasets tells a consistent story for the evolution of $M^\ast$ galaxy
progenitors. \S~9 summarizes our conclusions.    

All magnitudes here are relative to the AB system \citep{oke83}.  We
denote photometric magnitudes measured in the \hst/WFC3 F125W and
F160W passbands as $\wfcj$ and $\wfch$, respectively. Throughout, we
use $\ast$ in the subscript, $M_\ast$, to denote derived stellar
masses of individual galaxies.   We use $\ast$  in the superscript,
$M^\ast$, to denote the characteristic mass of the stellar mass
function.  For all derived quantities, where applicable we assume a
cosmology with $\Omega_m=0.27$, $\Omega_{\Lambda}=0.73$, and
$H_0=70.4$ km s$^{-1}$ Mpc$^{-1}$, \ed{consistent with the \textit{WMAP}
seven-year data \citep{koma11}}.  

\section{On the Properties of $M^\ast$ Galaxies: the MW and M31}\label{section:MW}

This paper focuses on the evolution of the main progenitors of
$M^\ast$ galaxies in two bins of stellar mass.   We define ``MW-mass''
and the ``M31-mass'' galaxies to be those galaxies with present-day
($z=0$) stellar masses near $M_\ast = 5 \times 10^{10}$~\msol\ and
$M_\ast = 10^{11}$~\msol, respectively.  These stellar masses are
consistent with the range for the MW and M31 currently published in
the literature \citep[see Mutch et al.\ 2011, references therein; and
also][]{mcmillan11,vandokkum13,licq14}, based on the modeling of the
MW and M31 luminosities with $M/L$ ratios consistent with that \ed{of}
a Chabrier 2003 initial mass function \citep[IMF;
see,][]{flynn06,geehan06}.  \citep[However, see the recent study
of][who derived a much lower mass for the MW compared to other
work.]{gibb14}.   As illustrated in figure~\ref{fig:mf}, the adopted
masses for the MW and M31 span the range  in the literature for
present-day ($z < 0.05$) values of $M^\ast$, which range from $6\times
10^{10}$ \citep{baldry08} to $9 \times 10^{10}$ \citep[accounting for
differences in the Hubble parameter and IMF]{bell03b,marc09}.
\ed{Therefore, our investigation probes the evolution of MW-mass and
M31-mass progenitors.  These bracket the observed range of $M^\ast$
galaxies, and allows us to compare the empirical evolution for such
galaxies that at present differ in stellar mass by a factor of two.}

Although throughout this paper we discuss the evolution of $M^\ast$
galaxies in subsamples of MW-mass and M31-mass galaxies, the MW and
M31 themselves may be outliers.  Indeed, there is growing evidence
that neither the MW nor M31  themselves are ``typical'' of the
\ed{galaxy population at these masses}. \citet{mutch11} presented a
comparison of the MW and M31 galaxies to other galaxies with similar
stellar masses selected from the Sloan Digital Sky Survey (SDSS).
They concluded that both the MW and M31 have bluer optical colors at
fixed stellar mass compared to galaxies matched in stellar mass and
morphology in SDSS: both the MW and M31 reside in the ``green valley''
of the galaxy color-mass distribution.   Mutch et al.\ concluded that
the MW and M31 are in the process of transitioning their global
properties from star-forming to more quiescent phases of galaxy
evolution.     In contrast, the ``typical'' $M^\ast$ galaxy is already
a red-sequence galaxy in SDSS. 

\ifsubmode
\begin{figure} 
\else
\begin{figure*}
\epsscale{1.} 
\fi 
\plotone{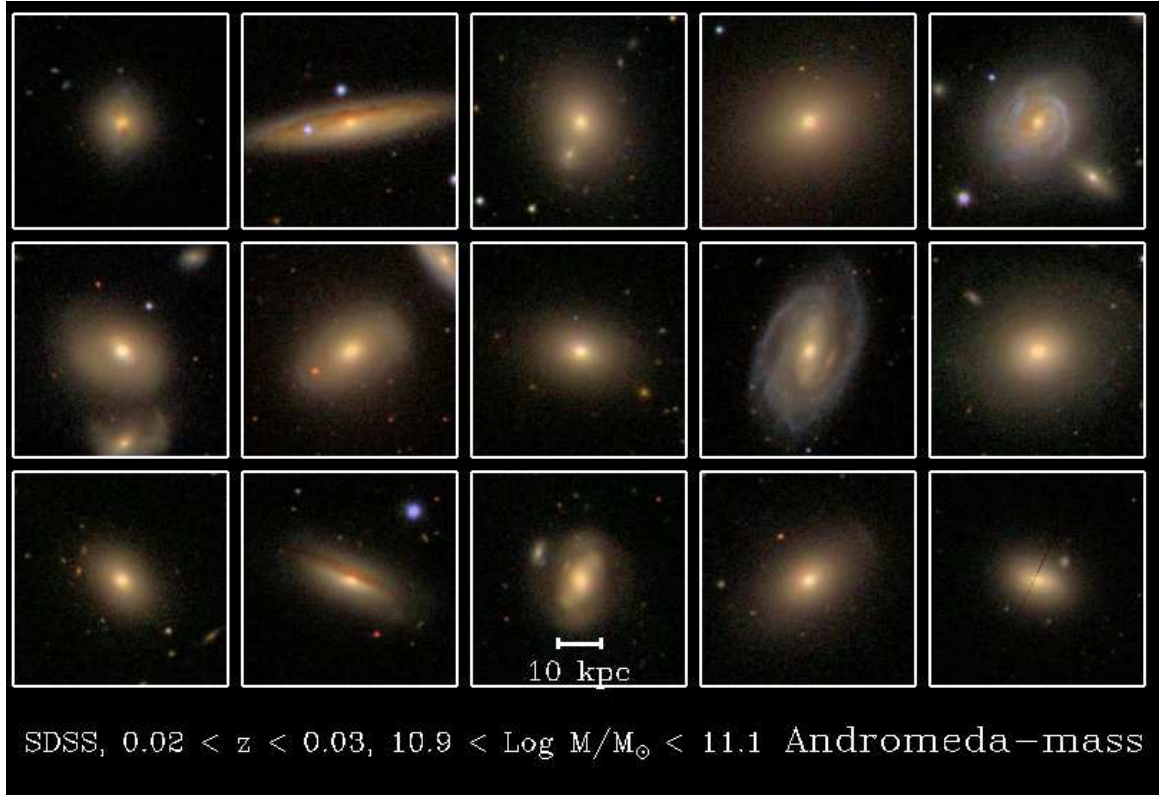}\caption{Montage of galaxies selected randomly
  from SDSS with $0.02 < z < 0.03$ and stellar mass $10.9 < \log
  M_\ast/\msol < 11.1$:  these are present-day M31-mass galaxies using our
  choice of stellar mass.   \ed{The images are SDSS $gri$-band composites}.  The montage shows that at $z\sim 0$ these galaxies 
  are dominantly spheroidal and early-type. Although
  some examples of disk galaxies with spiral-structures are
  evident, these structures are not the norm for M31-mass
  galaxies.    }\label{fig:sdssAndro}
\epsscale{1}
 \ifsubmode
\end{figure}
\else
\end{figure*}
\fi

\ifsubmode
\begin{figure} 
\else
\begin{figure*}
\epsscale{1.} 
\fi 
\plotone{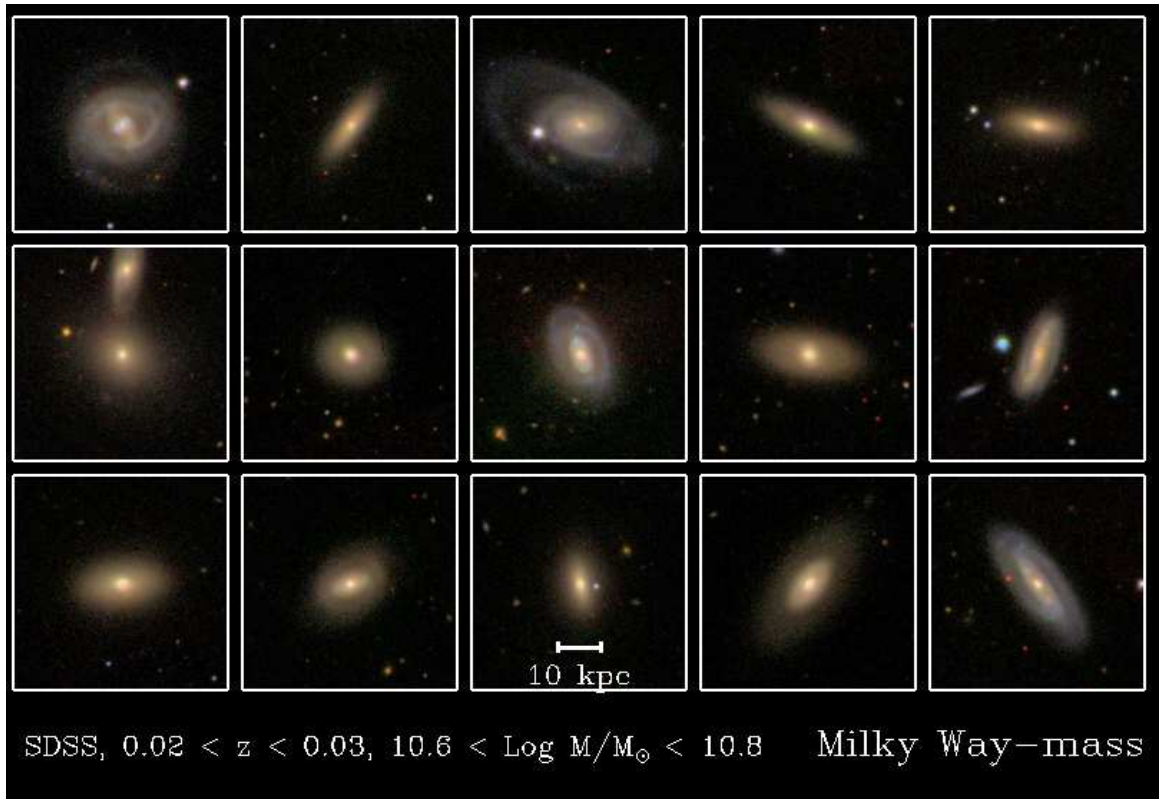}\caption{Montage of galaxies selected randomly from
SDSS with $0.02 < z < 0.03$ and stellar mass $10.6 < \log M_\ast/\msol <
10.8$:  these are the present-day MW-mass galaxies using our choice of
stellar mass.     \ed{The images are SDSS $gri$-band composites}. As with figure~\ref{fig:sdssAndro}, this montage shows
that galaxies at present with these stellar masses are dominantly
spheroidal and early-type, although some examples of disk galaxies
with blue (star-forming) spiral-structures are  present.    }\label{fig:sdssMW}
\epsscale{1}
 \ifsubmode
\end{figure}
\else
\end{figure*}
\fi

A perusal of M31- and MW-mass galaxies in the SDSS is consistent with
this conclusion.   Figures~\ref{fig:sdssAndro} and \ref{fig:sdssMW}
show montages of M31-mass and MW-mass galaxies randomly selected from
SDSS DR7 with $0.02 < z < 0.03$ and stellar mass $10.9 < \log
M_\ast/\msol < 11.1$, and $10.6 < \log M_\ast/\msol < 10.8$, within
0.1 dex of our adopted values for M31 and the MW, respectively
\citep[using stellar masses for SDSS DR7 derived from the MPA-JHU
value-enhanced
catalog,\footnote{{http://home.strw.leidenuniv.nl/$\sim$jarle/SDSS/}}][]{brin04}.
The montages in figures~\ref{fig:sdssAndro} and \ref{fig:sdssMW} show
that the typical M31-mass and MW-mass galaxies are spheroidal, or
reddened, bulge-dominated disks.  Qualitatively, many of these
galaxies appear more early-type in morphology compared to both the MW
and M31, except for a fraction of cases where bluer, spiral structures
are apparent.    

The preponderance of early-type morphologies among the MW-mass
galaxies is at odds with observations of the MW.    For example,
\citet{mutch11} argue that the MW is a Sb/c Hubble type.   The mass of
the MW's central supermassive black hole (SMBH) is low compared to
either its dark-matter halo, or its perceived bulge mass.  This may be
mitigated if the MW has only a pseudo-bulge \citep[where SMBH mass is
known to correlate with ``classical'' bulge mass,][]{korm11a}, and
these observations reinforce the idea that the morphology of the MW is
later type than the typical MW-mass galaxy in SDSS. 

Therefore, while both M31 and the MW are examples of $M^\ast$
galaxies, they are not themselves the most representative of the
$M^\ast$ population.      The results that we derive in this paper
pertain to the median evolution of \ed{galaxies with present-day
masses $5\times 10^{10}$~\msol\ and $10^{11}$~\msol}.  While this
provides insight into the formation and assembly history of the MW and
M31 themselves, it may be that these do not necessarily pertain to the
exact history for either galaxy.

\section{\zfourge\ and Ancillary Datasets}\label{section:zfourge}

The \zfourge\ survey (I.~Labb\'e et al.\ 2014, in preparation) is a deep
medium-band near-IR survey using the {\sc FourStar} instrument
\citep{pers13} mounted on the Magellan/Baade Telescope.    The main
\zfourge\ survey obtained very deep near-IR imaging in five adjacent
medium-band filters (\jone,\jtwo,\jthree,\hs,\hl) and a standard \ks\
filter.   The {\sc FourStar} \jone\ filter provides similar coverage
as the now more commonly used $Y$-band filter on near-IR imagers, and
the \jtwo\jthree\ and \hs\hl\ filter pairs divide the $J$-band and
$H$-band near-IR windows \citep[see, e.g.,][]{tilvi13}.  These
medium-band filters are very similar to those used by the NEWFIRM
Medium-Band survey \citep[NMBS,][]{vandokkum09a,whit11}, with small
differences \citep[particularly the central wavelength of the \jtwo\
filter, see][]{tilvi13}.  The filters provide $R$$\sim$10
``spectroscopy'' of the Balmer-break as it moves through these bands
at $1 < z < 4$.     As a result, the bands provide accurate
photometric redshifts $\sigma(z) / (1+z) \approx 1-2$\%
\ed{\citep[\eg,][T. Yuan et al.\ in
preparation]{vandokkum09a,whit11,spit12,kawi14}}.  

Here, we use the main \zfourge\ survey, which imaged three $11\arcmin
\times 11 \arcmin$ fields, widely separated on the sky:
the CDF-S, COSMOS, and UDS fields. The \zfourge\ pointings overlap
with the deepest portions of the CANDELS \hst\ imaging, and deep
\spitzer\ and \herschel\ imaging, described below.  Our {\sc FourStar}
images achieve depths of $\ks =$~24.80, 25.16, 24.63 AB mag, in each
field respectively ($5\sigma$) measured in $0\farcs6$-diameter
apertures, \ed{corrected to total apertures based on the
curve-of-growth for point sources}.  \ed{In addition, for the UDS field we
use a detection that is the sum of our {\sc FourStar} \ks\ image and
the \ks\ image from the UKIDSS
DR8\footnote{http://www.nottingham.ac.uk/astronomy/UDS/}.  The total
depth of this image is $\ks$=25.2 AB mag measured from the same
aperture as above}.   The depths in the other {\sc FourStar} bands are designed
to match the colors of red, passive galaxies at $z > 1$, reaching
$\jone \approx \ks + 1$~mag.    The data quality of the {\sc FourStar}
images is excellent, with the FWHM$\simeq$0.5--0.6\arcsec\ for the
point-spread function (PSF) for the stacked {\sc FourStar} images
\citep{tilvi13}.

We combined the {\sc FourStar} near-IR images with existing ancillary
ground-based imaging (spanning $U$ through $z$ bands), the CANDELS
\hst/ACS and WFC3 imaging \citep{grogin11,koek11}, and \spitzer/IRAC
imaging to generate multiwavelength catalogs spanning $0.3-8$~\micron\
\ed{\citep[the exact bands available depend on the field, see][and the
acquisition, data reduction, and description of the multiwavelength
catalogs will appear in C.\ Straatman et al.\ 2014, in
prep.]{tomc14}. \footnote{see also http://zfourge.tamu.edu}} For each
field, the ground-based and \hst\ images are convolved to match the
seeing in the image with the worst image quality (largest FWHM).
Photometry is measured in $1\farcs2$-diameter circular apertures, and
an aperture correction applied using the $\ks$ data for each source.
Typically, the relative flux for point sources between bands is
matched to better than 2\% for circular apertures with radii larger
than $0\farcs47$.  The IRAC 3.6, 4.5, 6.8, and 8.0~\micron\ data were
matched to the optical/near-IR catalogs using the procedure described
in \citet{labbe06,labbe10a}.

\ed{\subsection{Photometric Redshifts, Stellar Masses, \\ and Rest-Frame Colors}}

Photometric redshifts were derived using the full multiwavelength
catalogs spanning 0.3-8~\micron\ with EAZY \citep{bram08}.  \ed{EAZY
  reports small uncertainties on the photometric redshifts for the
  \zfourge\ samples.   For the $M^\ast$-progenitor subsamples used
  here, the average 68\%  uncertainties on the photometric redshifts range
  from $\sigma(z) /  (1+z) = 0.013 - 0.020$, \citep[see also discussion in][]{kawi14}. }
Rest-frame colors are derived using InterRest \citep{taylor09} using
the EAZY photometric redshifts.   \ed{We focus on the $U-V$ and
  $V-J$ rest-frame colors of the $M^\ast$ progenitor subsamples. We
  estimated uncertainties on these rest-frame colors, remeasuring the colors
  in a monte carlo  simulation, perturbing the fluxes of each object
  1000 times and taking the inter-68\%-tile range as the uncertainty.   The
  average uncertainties on these rest-frame colors are $\sigma(U-V) =
  0.06 - 0.12$~mag and $\sigma(V-J)=0.10-0.19$~mag for the $M^\ast$
  progenitors over the redshift range $z\sim 0.5 - 3$.}

Stellar masses were derived by fitting \citet{bruz03} stellar
population synthesis models with FAST \citep{kriek09b} using a
\citet{chab03} IMF, \ed{solar metallicity, and using exponentially
declining star-forming histories ($\Psi$ $\sim \exp(-t/\tau)$), where
the age ranges from $\log t/\mathrm{yr}$ = 7.5 -- 10.1 in steps of 0.1
dex and the e-folding timescale ranges from $\log \tau/\mathrm{yr}$ =
7.0--11.0 in steps of 0.2 dex.   The effects of dust attenuation were
included using the prescription from \citet{calz00} ranging from
$A_V$=0--4.0 mag in steps of 0.1 mag.    \ed{Adopting different
exctinction laws can affect the stellar masses at the $\sim$0.2-0.3
dex level \citep[\eg,][]{papo01,marc09,tilvi13}.}  While we expect the
metallicity of the $M^\ast$ progenitors to evolve over the redshift
range studied here, assuming different metallicities in the fitting of
the spectral-energy distributions has only a minor impact on stellar
masses \citep[e.g.,][]{papo01,gall09,marc09}.  Similarly, using
different star-formation histories (e.g., models with SFRs that
increase with time; ``delayed'' exponentially decling models) can
introduce systematic uncertainties at the $\simeq$0.2 dex level
\citep[see, e.g.,][]{mara10,kslee10,papo11}.  The typical statistical
uncertaintes on the stellar masses  from FAST for the $M^\ast$
progenitors are formally $0.10-0.14$~dex depending on mass and
redshift.  Therefore, we expect the combined uncertainties on the stellar
masses (statistical and systematic) to be $<$0.2-0.3 dex level (factor
of 2), dominated by systematics.}

\ed{\subsection{Stellar Mass Completeness}}

\ed{We estimated the completeness in the current \zfourge\
images and catalogs, and in our samples of $M^\ast$ galaxies (defined
in \S~5) in two ways.  First, we compared the completeness in
stellar mass in the \zfourge\ catalogs to the catalogs from 3D-HST
\citep[see below]{skel14}, which provide an empirical test of our
catalogs to $z \lsim 3$ where 3D-HST achieves deeper stellar mass
completeness.  Second, we performed simulations where we
inserted fake point sources in the \ks-detection image for each of the
three \zfourge\ fields.   We allow the sources to have magnitudes
chosen from a wide distribution, and we allow the sources to be
located anywhere in the detection image.  In this way random objects
may fall within the isophote of real objects in the image, and
therefore our completeness simulations include the effects from
blended objects.  We measure the 80\% completeness limit to be the
magnitude where we recover 80\% of the fake sources using the same
detection parameters as for the real catalog.  For the \zfourge\
CDF-S, COSMOS, and UDS catalogs the 80\% completeness limits are $\ks
= 24.53$, 24.74, and 25.07 AB mag, respectively (the 90\% completeness
limits are approximately 0.2 mag shallower in each field).   From our
simulations, we also estimate that blended objects account for 5\% of
this incompleteness.  For the remainder of this work, we consider
samples where the data are formally 80\% complete.  }  

\ed{The 3D-HST catalogs provide a estimate of our stellar mass
completeness for $z < 3$ because at these redshifts the (deeper,
\wfch-band selected) 3D-HST catalog achieves a lower stellar-mass
limit than our (shallower, \ks-band-selected) \zfourge\ catalog.  We
matched sources in \zfourge\ to 3D-HST in the regions where they
overlap, and we computed the completeness as the fraction of sources
in 3D-HST detected in \zfourge\ in bins of stellar mass and redshift.
The 80\% completeness in stellar mass is} \edtoo{$\log M/\msol$ = 8.8}
\ed{, 9.2, 9.4, 9.5, and 9.8~dex  in bins of $1 < z < 1.5$, $1.5 < z <
2$, $2 < z < 2.5$, $2.25 < z < 2.75$, and $2.5 < z < 3$, respectively
(where the penultimate bin is about the same redshift range as the
highest redshift bin for our MW progenitor subsample).  This test also
accounts for completeness effects as a result of galaxy properties
themselves, including blending between sources that are resolved in
the \hst\ catalog, but blended at the {\sc FourStar} resolution, the
intrinsic colors of galaxies (including possible dust-obscured, low
mass galaxies),  and for the fact that the galaxies in our samples are
not point sources. }

\ed{Based on the comparison to 3D-HST and the point-source simulations,
the MW-mass progenitors are $>$90\% complete for $z < 2.2$.  At this
redshift, the MW-mass progenitors are already mostly star-forming,
with blue colors and low dust obscuration (based on their
$\lir/L_\mathrm{UV}$ ratios, see \S~5 and \S~7, below).   Such blue
objects have lower $M/L$ ratios, and are complete to lower stellar
mass than the completeness derived for the \ks-band limit.    Because
the  MW-mass progenitors are already blue with no indication of a
significant population of very dust-reddened or quiescent progenitors,
it seems unlikely that such a population would suddenly be part of the
MW-mass progenitor population at higher redshift at lower stellar
masses.  Therefore, we expect the MW-mass progenitors to be reasonably
(80\%) complete  in their highest redshift bin, $2.2 < z < 2.8$, and
this is consistent with the comparison to 3D-HST.}

\ed{The M31-mass progenitors are $>90$\% complete for $z < 2.8$.  The
formal 80\% completeness stellar-mass limit (from our simulations and
the \ks-band limit) is moving through the highest-redshift bin for the
M31-mass progenitors, $2.8 < z < 3.5$, but we expect higher
completeness because the populations have relatively blue colors at
lower redshifts, $z < 2.8$.  Nevertheless, the 
stellar-mass completeness values are only estimates, and these would
be biased if there exists a significant, undetected population of
low-mass, dusty or quiescent red galaxies.  Any conclusions about the
$M^\ast$ galaxies in their highest redshift bins could be biased if
these samples are missing a hypothetical population of redder galaxies
than those counted in our simulations. }

\subsection{\hst\ imaging}

The three \zfourge\ fields (COSMOS, CDF--S, UDS) overlap with the
CANDELS \hst\ imaging with WFC3 using the F125W and F160W passbands.
The \hst\ data provide higher angular resolution imaging
\ed{\citep[FWHM$\simeq 0\farcs2$, see][]{koek11}} compared to any of
the ground-based datasets, and this allows us to resolve structures
down to $\sim$1~kpc.     We make use of the galaxy structural
properties (effective sizes and \sersic\ indices) measured with the
CANDELS \hst\ imaging with WFC3 published by \citet{vanderwel12}.
Throughout this work we focus on the sizes and \sersic\ indices
measured in the F160W band as this allows measurements in the
rest-frame 4000~\AA\ (approximately the $B$-band) out to  $z\sim 3$.
In addition, the CANDELS coverage of the CDF-S field includes F105W
imaging, as well as the ACS imaging from $0.4-1$~\micron\ in the
F435W, F606W, F775W, and F850LP bandpasses from GOODS \citep{giav04a}.
At lower redshifts, the F160W band probes light from longer rest-frame
wavelengths.  However, our tests using data from the WFC3 F105W
passband in the CDF-S show that the differences in the structural
parameters are minor, and that none of our conclusions would be
affected.  

We matched the sources in the \zfourge\ catalogs to those in
\citet{vanderwel12} using a matching radius of $0\farcs5$.  We then
adopt the effective semimajor axis and \sersic\ index for each source
from the van der Wel et al.\ catalog.   Here, the effective sizes we
report are the circularized effective radius, $\reff = \sqrt{ab} =
a_\mathrm{eff} \sqrt{q}$, where $a_\mathrm{eff}$ is the effective
semimajor axis measured in van der Wel et al., and $q = b/a$ is the
ratio of the semiminor to semimajor axes.  

\subsection{\spitzer\ and \herschel\ Far-IR Imaging}

The \zfourge\ fields cover areas with imaging from \spitzer/MIPS and
\herschel/PACS.   We use the deepest of these data to measure the
mid-IR and far-IR emission for galaxy populations selected from
\zfourge.     In practice, we are interested here in the average IR
emission from galaxies in our samples.  To ensure we are not biased by
the subset of galaxies detected in the mid-IR and far-IR data, we will
stack the IR data at the locations of the galaxies in our samples to
produce average flux density measurements (see \S~\ref{section:ir}). 

For the \zfourge\ CDF-S field, we used \spitzer/MIPS 24~\micron\
imaging from the GOODS \spitzer\ Legacy program \citep[PI:
M.~Dickinson, see also][]{magn11}.  For the \herschel/PACS 100 and 160~\micron\
imaging we used the data taken by the GOODS-Herschel survey
\citep[GOODS-H, a \herschel\ Key Project,][]{elbaz11}. 

For the COSMOS field, we used MIPS 24~\micron\
imaging from the SCOSMOS \spitzer\ Legacy program (PI: D.\
Sanders)\footnote{http://irsa.ipac.caltech.edu/data/SPITZER/S-COSMOS}.
We also \ed{used} deep PACS 100 and 160~\micron\
data from CANDELS-Herschel (CANDELS-H, H.\ Inami, et al.,  in prep.),
reduced in the same \ed{way} as GOODS-H.      

For the UDS field, we used the MIPS 24~\micron\ imaging
from SpUDS \spitzer\ Legacy program (PI:
J. Dunlop)\footnote{http://irsa.ipac.caltech.edu/data/SPITZER/SpUDS},
combined with deep data taken with PACS at 100 and
160~\micron\ also as part of CANDELS-H.

\section{Selecting the Progenitors of $M^\ast$ Galaxies}\label{section:samples}

\ed{There is a growing body of works in the literature that select the
  progenitors of galaxies at higher redshifts (earlier cosmic epochs)
  by requiring that they have the same co-moving, cumulative number density
\citep{brown07,brown08,cool08,vandokkum10,vandokkum13,papo11,beza11,
  bram11,fuma12,cons13,leja13,muzz13c,patel13a,lund14,tal14,marc14}.}  This
method is an approximation, as it neglects variations (scatter) in mass growth,
including effects of galaxy mergers on the mass-rank order of
galaxies.  \citet{leja13} compared the selection of progenitors using
constant number density to other means using a mock catalog from the
Millennium simulation.   They showed that selecting galaxies based on
constant number density reproduces the stellar mass of progenitors,
but with uncertaintites of 0.15 dex from $z=3$ to 0.  \citet{behr13c} recently
discussed how the selection at constant number density ignores scatter
in mass accretion histories and mergers, which can lead to errors in
the mass evolution of galaxy progenitors on the order of $d(\log M_\ast)/dz
= 0.16$ dex (i.e., factor of $\approx$40\% per unit redshift).   \ed{This
error is exacerbated for galaxies with lower $z=0$ stellar masses
(larger number densities).}

Here,  we have used results of a multi-epoch abundance matching (MEAM)
method  \citep{moster13} to identify the main progenitors of
present-day $M^\ast$ galaxies at higher redshifts.  Moster et al.\
derived a redshift-dependent parametrization of the stellar-mass to
halo-mass relation, whereby they populate dark-matter halos and
subhalos in the Millennium simulations with galaxies that follow a
distribution of stellar mass, such that the evolution of observed
stellar mass functions are reproduced simultaneously.
\ed{\citet{behr13b}} used a similar method (also called
``stellar-halo-mass abundance matching'') applied to the independent
Bolshoi simulation to show that this reproduces both the stellar-mass
function evolution and the star-formation history over a large range
of galaxy mass and redshift ($0 < z < 8$).  Because the
abundance--matching methods of Moster et al.\ and Behroozi et al.\
track the evolution of galaxies with their dark--matter halo
evolution, they naturally correct for variations in galaxy mass growth
and galaxy mergers compared to techniques that select progenitors at
constant number density.  \ed{Nevertheless, as figure~\ref{fig:ndense}
shows, all methods produce very similar mass evolution \citep[see
also,][]{leja13}.}

\ifsubmode
\begin{figure} 
\epsscale{0.75}
\else
\begin{figure}  
\epsscale{1.15} \fi
\plotone{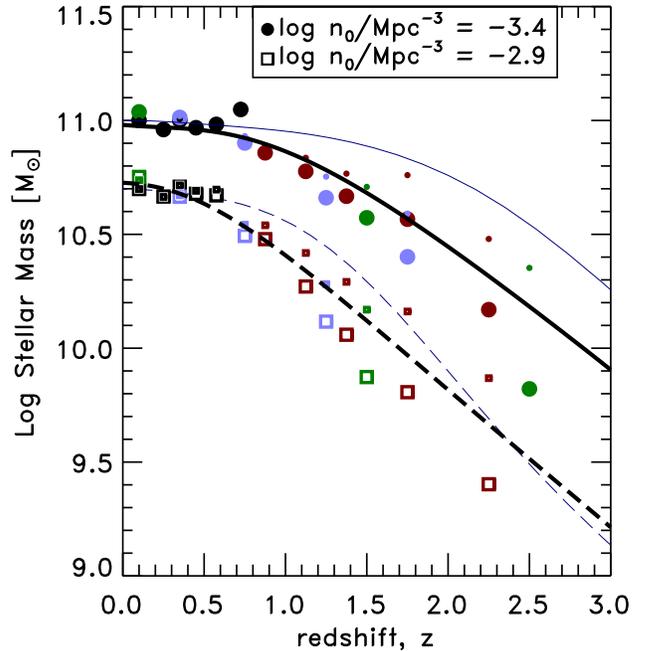}
\caption{Stellar-mass evolution of galaxies of M31-mass (present-day
stellar mass $10^{11}$~\msol) and MW-mass (stellar mass $5\times
10^{10}$~\msol)  progenitors as a function of redshift.  \ed{The data
points show the stellar-mass evolution of galaxies selected by their
number density for present-day ($z$=0) values $\log
(n_0/\mathrm{Mpc}^{-3}) = -2.9$ for the MW-mass (open boxes)  and
$\log(n_0/\mathrm{Mpc}^{-3}) = -3.4$ for the M31-mass (filled circles)
progneitors.      The different colors represent values for different
literature mass functions (black: Moustakas et al.\ 2013;  blue:
Muzzin et al.\ 2013; green: Marchesini et al.\ 2009; red: Tomczak et
al.\ 2013), \edtoo{where we show points only at redshifts where the mass
functions are complete.} The small data points show the evolution for constant
co-moving number density,  derived by integrating stellar mass
functions down to the same number density at each redshift.  The large
data points show the mass evolution for an evolving number density
from \citet{behr13c}.}  The \ed{thick solid and dashed-line curves
show the stellar-mass evolution from the abundance-matching model of
\citet{moster13} for galaxies with $M_\ast=10^{11}$ and
$M_\ast=5\times 10^{10}$~\msol, respectively, at $z=0$.  Galaxies with
these stellar masses at $z=0$ have halo masses of  $M_h = 10^{13}$ and
$2.5\times 10^{12}$ \msol, respectively in this model.   The thin
solid and dashed lines show the evolution for galaxies with the same
present-day stellar masses based on modeling their median star-formation
histories \citep{behr13b}.}  Here we use the stellar-mass evolution
from Moster et al.\ model to select progenitors of MW-mass and
M31-mass galaxies.  }\label{fig:ndense}    \epsscale{1} 
\ifsubmode
\end{figure}
\else
\end{figure}
\fi

We derive the stellar-mass evolution of galaxy progenitors using the
results of \citet{moster13}, who provided fitting functions for the
star-formation history and mass-accretion history for galaxies of
arbitrary present-day stellar mass.   We integrated the
\citet{moster13} fitting functions with respect to time, accounting
for mass losses from stellar evolution \citep[see,][their eqn.\
16]{moster13} to derive the conditional stellar-mass evolution of
galaxies.   Figure~\ref{fig:ndense} shows the stellar-mass evolution
of present-day galaxies with $5\times 10^{10}$~\msol\ (MW-mass
galaxies) and $10^{11}$~\msol\ (M31-mass galaxies).\footnote{The
stellar-mass evolution we derive via integrating the star-formation
and accretion histories matches the direct results from Moster et al.\
at $z\sim 0$, but produces masses $\lsim$0.15 dex lower at $z=2$ (B.\
Moster 2013, private communication).  These are both within the
plausible range of mass-growth histories in Moster et al.\ (2013), and
so both are equally consistent. }
This growth is more rapid at $z > 1$, with $\log M_\ast \propto -1.1
\Delta z$, which can be compared to the predicted halo growth based on simple
theoretical grounds, where $\log M_h \propto -0.8 \Delta z$ \citep[valid at
$z>1$]{dekel13}.  \ed{This is expected as at these redshifts the
halo mass corresponding to the peak value in $M_\ast/M_h$ (related to
the star-formation efficiency) decreases with redshift \citep[\eg,][]{behr10, behr13b}. }

\ed{Figure~\ref{fig:ndense} shows the expected stellar-mass evolution at
constant \ed{and evolving number density \citep[using the prescription
of][]{behr13c}}.  Using the stellar--mass function at $z\sim 0.1$ from
SDSS \citep{mous13} we find that galaxies with present day stellar
masses of $5\times 10^{10}$~\msol\ and $10^{11}$~\msol\ have number
densities $\log (n/\mathrm{Mpc}^{-3}) = -2.9$ and -3.4, respectively,
where rarer objects with lower number density have higher mass.  We
then integrate the literature mass functions
\citep{mous13,muzz13c,marc09,tomc14}  to the appropriate number
density at different redshifts down to the stellar mass such that
$n(>M_\ast)$ = constant (for constant number density) or to the
evolving number density predicted by \citet{behr13c}.} \edtoo{We only
include data points in figure~\ref{fig:ndense} where the mass
functions are complete.}

A comparison of the data points and curves in fig.~\ref{fig:ndense}
shows that for $M^\ast$-mass galaxies the stellar-mass evolution
derived using the \citet{moster13} abundance matching is mostly
consistent to that measured using samples at fixed number density.
There is a slight bias in the stellar-mass evolution at constant
number density toward higher masses at higher redshift. For example,
the evolution at constant number density from the Tomczak et al.\
(2013) mass function gives masses larger by $\simeq$0.1--0.2 dex at
$z$$>$2 for the MW and M31-mass progenitors compared to the
abundance-matching results.  This is qualitatively consistent with the
findings of \citet{leja13} and \citet{behr13c}, both of whom find that
the number density of galaxy progenitors at higher redshifts shifts to
higher values, implying they correspond to lower stellar masses
compared to a constant number density selection.  The effect is about
$0.1$~dex from $z$=0 to 3 \citep{leja13}, which  is consistent with
our observed trend.  

\ed{\citet{behr13c} provide the number density evolution of the
progenitors of a present-day galaxy population with some ($z=0$)
number density. Figure~\ref{fig:ndense} shows this mass evolution
using the median number density evolution from Behroozi et al.\ with
the same literature stellar-mass functions.   The evolving number
density predicts lower stellar masses compared to the constant number
density.     The truth is probably in between these as the evolving
number density predictions assume a dark matter merger rate which may
not track exactly the galaxy merger rate.  In many cases, the evolving
number density also predicts lower stellar masses compared to either
the \citet{moster13} and \citet{behr13b} models.  We attribute this to
uncertainties in the observed stellar mass functions at the low-mass
end, where small uncertainties in the number densities lead to large
uncertainties in the stellar mass evolution.  }

Therefore,  here we will use the stellar-mass evolution predicted by
the abundance matching technique of \citet{moster13} to select
progenitors of M31 and MW-mass galaxies.  The evolution predicted by
\citet{moster13} is nearly identical to that of \ed{\citet[as illustrated
in figure~\ref{fig:ndense}]{behr13b}}, where
the latter used a simultaneous fit to the stellar-mass functions,
specific star-formation rates, and cosmic star-formation rates.  There
is negligible difference in the evolution of the MW progenitors
between the two models.  The biggest difference is for the M31-mass
progenitors (with $z=0$ stellar mass, $10^{11}$~\msol), where the
results of \citet{behr13b} predict higher stellar masses than those of
\citet{moster13} with a difference that increases with redshift up to
0.3 dex (factor of $\sim$2) at $z=3$.    \ed{Because we select
progenitors in bins of $\pm$0.25 dex about the median mass,  our
results would not change significantly if we used the latter instead.
The Moster et al.\ model predicts a smaller difference in stellar mass
between the M31 and MW progenitors at fixed redshift, and therefore
our conclusions are, if anything, conservative in that any differences
in the populations would presumably be accentuated using the Behroozi
et al.\ model.}

\ifsubmode
\begin{figure} 
\epsscale{1.1}
\else
\begin{figure*}  
\epsscale{1.15}
\fi 
\plottwo{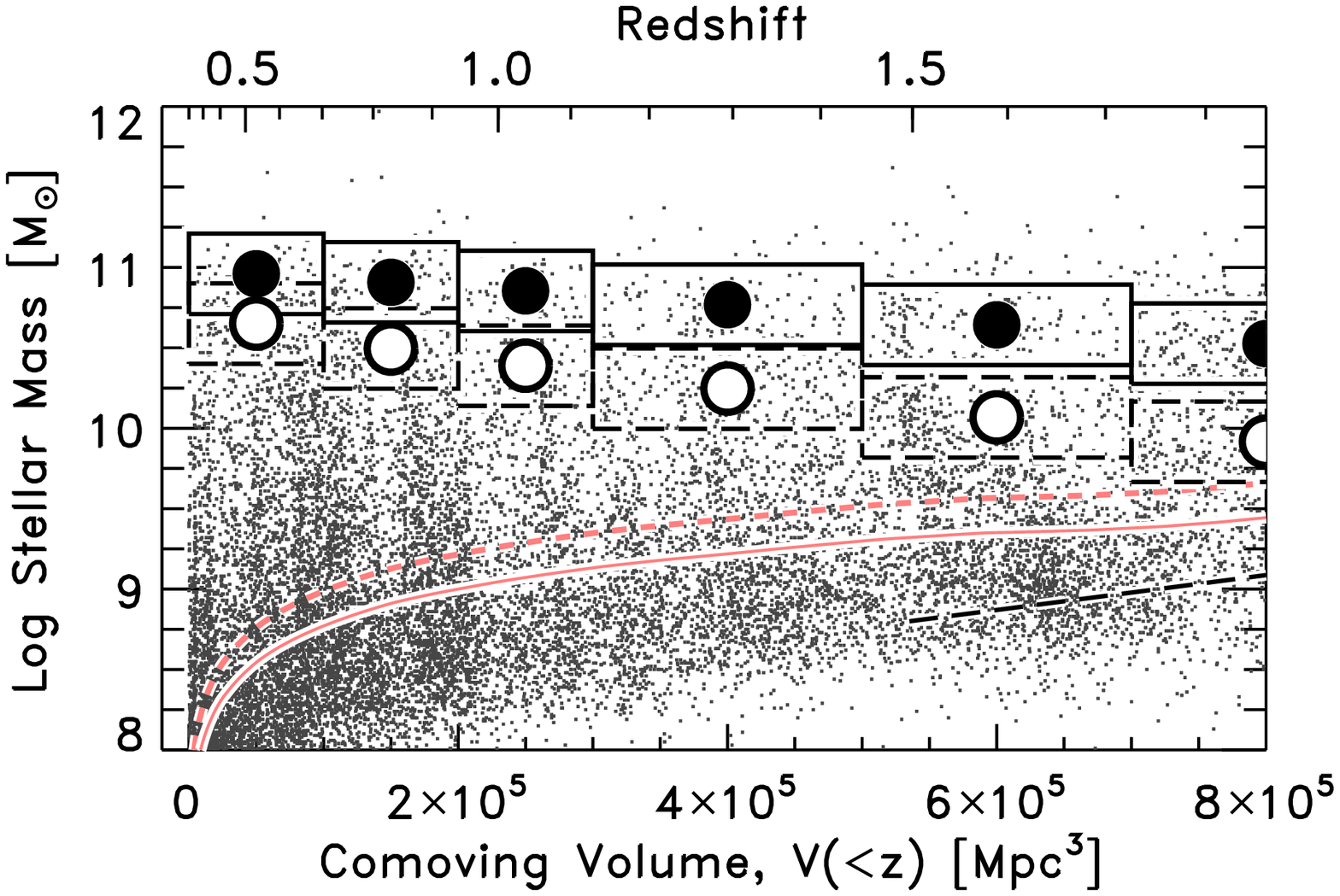}{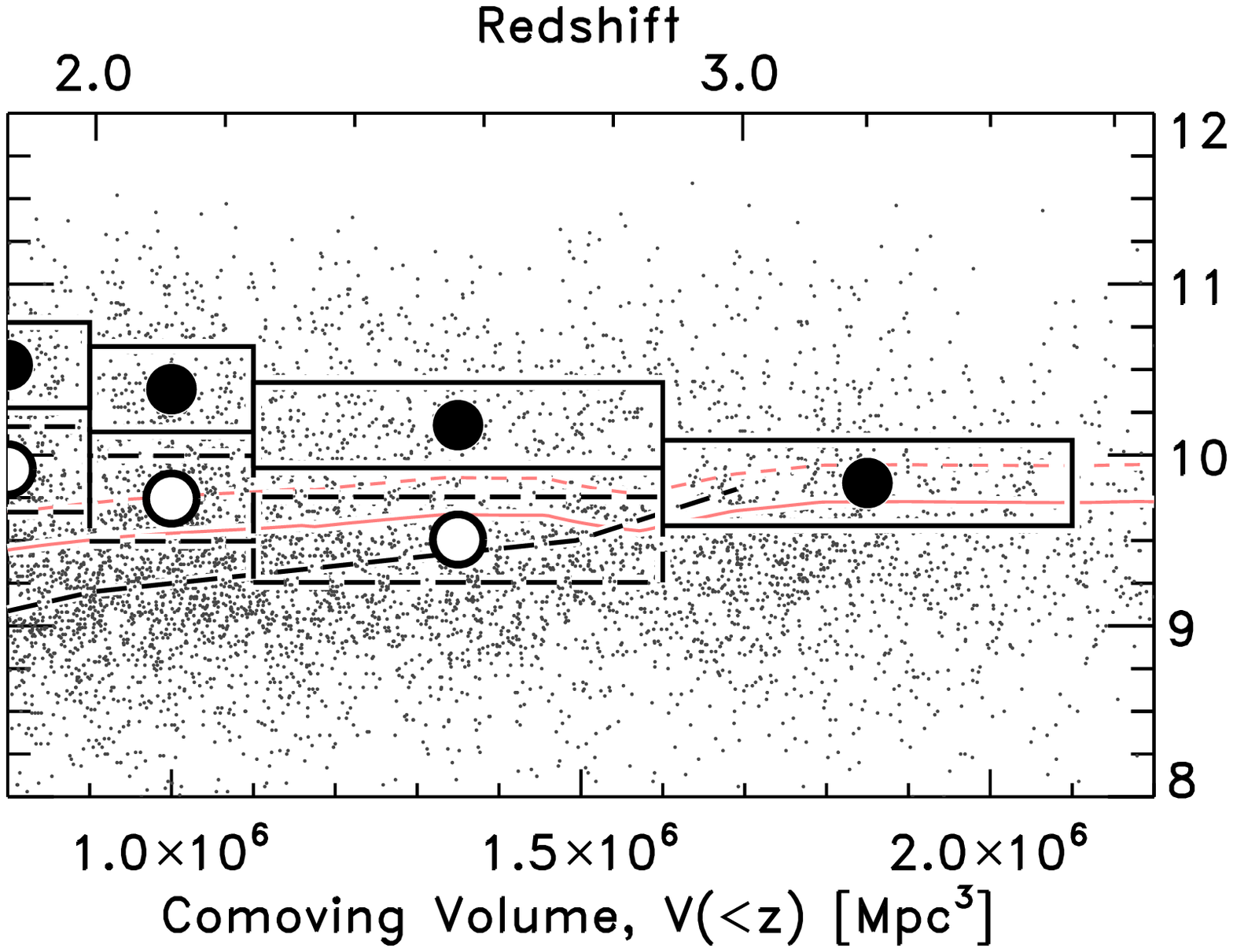}
\caption{Selection of $M^\ast$ progenitor galaxies in \zfourge.  The
data points show the stellar masses of all galaxies in the \zfourge\
COSMOS, CDF-S, and UDS fields as a function of comoving volume within
each redshift.  The scale of the abscissa changes between the left and
right panels for clarity.   The large circles indicate the central
stellar mass value in bins of comoving volume of the M31-mass (filled
circles) and MW-mass (open circles) progenitors selected from the
abundance matching of \citet{moster13} as described in the text.  The
solid-line and dashed-line boxes show the bins in comoving-volume and
stellar-mass used to select each progenitor subsample for the M31-mass
and MW-mass progenitors, respectively.   The volume bin width
increases at higher redshift as a trade-off between volume and
lookback time.  The red curves show the stellar-mass completeness limit for
red, passive galaxies defined as a stellar population formed at $z_f
= 5$ with no subsequent star-formation and no dust extinction for the
\ed{\ks-band\ limits derived from simulations for the CDF-S (dashed curve)
and UDS (solid curve).  \edtoo{The black dashed line shows the 80\%
completeness limit derived from the comparison to 3D-HST for galaxies
at $1.5 <z<3$.}}  }\label{fig:massvol}
\epsscale{1} 
\ifsubmode
\end{figure}
\else
\end{figure*}
\fi

\ifsubmode 
\begin{deluxetable}{lccccccccccc}
\tabletypesize{\scriptsize}
\else
\begin{deluxetable*}{lccccccccccc}
\fi
\tablecaption{Properties of $M^\ast$ Galaxy Properties\label{table:main}}
\tablecolumns{12}
\tablewidth{0pc}
\tablehead{
\colhead{redshift} &
\colhead{} &
\colhead{Median} &
\multicolumn{3}{c}{Number of objects per field} & 
\colhead{$r_\mathrm{eff}$} &
\colhead{} &
\colhead{$U-V$} &
\colhead{$V-J$} &
\colhead{$L_{2800}$} & 
\colhead{quiescent} \\
\colhead{range} & 
\colhead{$\log M_\ast/\msol$} & 
\colhead{$\log M_\ast/\msol$} & 
\colhead{CDFS} & 
\colhead{COSMOS} & 
\colhead{UDS} & 
\colhead{(kpc)} & 
\colhead{$n$} & 
\colhead{(mag)} & 
\colhead{(mag)} & 
\colhead{($10^{9}\ \lsol$)} & 
\colhead{fraction} \\
\colhead{(1)} & 
\colhead{(2)} & 
\colhead{(3)} & 
\colhead{(4)} & 
\colhead{(5)} & 
\colhead{(6)} & 
\colhead{(7)} & 
\colhead{(8)} & 
\colhead{(9)} & 
\colhead{(10)} &
\colhead{(11)} &
\colhead{(12)} }
\startdata
\multicolumn{12}{c}{Andromeda-like Progenitors} \\[3pt]
0.2 < z < 0.7 &     10.96 & 10.85 & \phn 31 & \phn 18 & \phn 20 & $ 3.6^{+1.3}_{-1.0} $ & $ 4.2^{+1.3}_{-1.5} $ & $ 2.0^{+0.2}_{-0.2} $ & $ 1.3^{+0.1}_{-0.1} $ & \phn $ 1.7^{+1.4}_{-0.5} $ & $ 0.85 \pm 0.04 $  \\
0.7 < z < 0.9 &     10.91 & 10.81 & \phn 39 & \phn 11 & \phn 19 & $ 3.0^{+1.8}_{-1.1} $ & $ 3.6^{+1.3}_{-1.0} $ & $ 1.9^{+0.2}_{-0.3} $ & $ 1.3^{+0.2}_{-0.2} $ & \phn $ 2.0^{+1.9}_{-0.7} $ & $ 0.70 \pm 0.05 $  \\
0.9 < z < 1.1 &     10.85 & 10.80 & \phn 15 & \phn 26 & \phn 23 & $ 3.2^{+1.4}_{-1.3} $ & $ 3.0^{+1.2}_{-1.3} $ & $ 1.7^{+0.2}_{-0.3} $ & $ 1.4^{+0.2}_{-0.2} $ & \phn $ 4.2^{+4.3}_{-1.1} $ & $ 0.47 \pm 0.06 $  \\
1.1 < z < 1.4 &     10.77 & 10.70 & \phn 39 & \phn 23 & \phn 44 & $ 2.3^{+1.3}_{-1.3} $ & $ 2.5^{+2.5}_{-1.3} $ & $ 1.7^{+0.2}_{-0.4} $ & $ 1.2^{+0.3}_{-0.2} $ & \phn $ 4.4^{+4.5}_{-1.9} $ & $ 0.60 \pm 0.04 $  \\
1.4 < z < 1.7 &     10.64 & 10.62 & \phn 36 & \phn 40 & \phn 63 & $ 1.7^{+1.7}_{-0.9} $ & $ 2.2^{+1.7}_{-1.3} $ & $ 1.6^{+0.3}_{-0.3} $ & $ 1.3^{+0.4}_{-0.2} $ & \phn $ 5.1^{+5.2}_{-2.3} $ & $ 0.47 \pm 0.04 $  \\
1.7 < z < 2.0 &     10.53 & 10.48 & \phn 59 & \phn 34 & \phn 38 & $ 2.1^{+1.9}_{-1.2} $ & $ 1.8^{+1.7}_{-1.1} $ & $ 1.5^{+0.3}_{-0.3} $ & $ 1.3^{+0.4}_{-0.3} $ & \phn $ 4.5^{+5.6}_{-2.2} $ & $ 0.33 \pm 0.04 $  \\
2.0 < z < 2.2 &     10.38 & 10.36 & \phn 51 & \phn 29 & \phn 43 & $ 2.4^{+1.4}_{-1.4} $ & $ 1.0^{+1.5}_{-0.5} $ & $ 1.2^{+0.5}_{-0.4} $ & $ 1.1^{+0.5}_{-0.4} $ & \phn $ 7.3^{+5.7}_{-4.5} $ & $ 0.31 \pm 0.03 $  \\
2.2 < z < 2.8 &     10.17 & 10.15 & \phn 57 & \phn 67 & \phn 86 & $ 2.1^{+1.1}_{-0.9} $ & $ 1.1^{+2.0}_{-0.6} $ & $ 0.9^{+0.6}_{-0.3} $ & $ 0.8^{+0.5}_{-0.4} $ &     $ 13.5^{+11.9}_{- 7.8} $ & $ 0.13 \pm 0.02 $  \\
2.8 < z < 3.5 & \phn 9.84 & 9.80 & \phn 95 & \phn 72 & \phn 77 & $ 1.2^{+0.9}_{-0.4} $ & $ 1.3^{+1.9}_{-0.7} $ & $ 0.6^{+0.4}_{-0.3} $ & $ 0.3^{+0.9}_{-0.6} $ &     $ 20.6^{+ 9.9}_{- 8.3} $ & $ 0.04 \pm 0.01 $  \\\hline
\\ \multicolumn{12}{c}{MW-like Progenitors} \\[3pt]
0.2 < z < 0.7 &     10.65 & 10.60 & \phn 59 & \phn 45 & \phn 47 & $ 2.6^{+1.5}_{-1.1} $ & $ 3.4^{+1.7}_{-1.7} $ & $ 1.9^{+0.2}_{-0.3} $ & $ 1.3^{+0.2}_{-0.1} $ & \phn $ 1.3^{+1.4}_{-0.4} $ & $ 0.74 \pm 0.03 $  \\
0.7 < z < 0.9 &     10.50 & 10.47 & \phn 81 & \phn 36 & \phn 43 & $ 2.1^{+1.4}_{-1.0} $ & $ 2.7^{+1.4}_{-1.4} $ & $ 1.7^{+0.2}_{-0.3} $ & $ 1.3^{+0.3}_{-0.2} $ & \phn $ 1.5^{+1.6}_{-0.6} $ & $ 0.54 \pm 0.03 $  \\
0.9 < z < 1.1 &     10.39 & 10.35 & \phn 44 & \phn 43 & \phn 35 & $ 2.3^{+2.2}_{-1.3} $ & $ 2.1^{+1.5}_{-1.2} $ & $ 1.6^{+0.3}_{-0.3} $ & $ 1.2^{+0.4}_{-0.2} $ & \phn $ 2.3^{+3.1}_{-0.9} $ & $ 0.41 \pm 0.04 $  \\
1.1 < z < 1.4 &     10.25 & 10.21 & \phn 78 & \phn 69 & \phn 81 & $ 2.2^{+1.6}_{-1.1} $ & $ 1.5^{+2.1}_{-0.8} $ & $ 1.4^{+0.4}_{-0.4} $ & $ 1.2^{+0.4}_{-0.3} $ & \phn $ 3.4^{+5.4}_{-1.9} $ & $ 0.29 \pm 0.03 $  \\
1.4 < z < 1.7 &     10.07 & 10.06 & \phn 82 & \phn 67 & \phn 99 & $ 2.1^{+1.3}_{-1.0} $ & $ 1.2^{+1.5}_{-0.6} $ & $ 1.1^{+0.5}_{-0.3} $ & $ 1.0^{+0.4}_{-0.3} $ & \phn $ 5.5^{+6.1}_{-3.6} $ & $ 0.20 \pm 0.02 $  \\
1.7 < z < 2.0 & \phn 9.92 & 9.88 & \phn 84 & \phn 51 & \phn 75 & $ 2.2^{+1.2}_{-0.9} $ & $ 1.1^{+1.1}_{-0.6} $ & $ 0.9^{+0.4}_{-0.3} $ & $ 0.8^{+0.5}_{-0.4} $ & \phn $ 8.9^{+9.2}_{-5.3} $ & $ 0.12 \pm 0.02 $  \\
2.0 < z < 2.2 & \phn 9.75 & 9.70 &     102 & \phn 70 & \phn 93 & $ 1.7^{+0.8}_{-0.6} $ & $ 1.3^{+1.4}_{-0.6} $ & $ 0.6^{+0.4}_{-0.3} $ & $ 0.5^{+0.4}_{-0.3} $ &     $ 11.7^{+ 5.5}_{- 5.0} $ & $ 0.06 \pm 0.01 $  \\
2.2 < z < 2.8 & \phn 9.51 & 9.48 &     173 &     197 &     224 & $ 1.3^{+0.8}_{-0.5} $ & $ 1.3^{+1.4}_{-0.7} $ & $ 0.6^{+0.3}_{-0.3} $ & $ 0.3^{+0.5}_{-0.4} $ &     $ 13.0^{+ 6.4}_{- 4.8} $ & $ 0.03 \pm 0.01 $  \\\hline
\enddata
\tablecomments{(1) Redshift range of bin, \ed{(2) central value of the stellar mass used to select progenitors in the redshift bin; galaxies are selected within $\pm$0.25 dex of this value in this bin.} \ed{ (3) median stellar mass of selected galaxies in the redshift bin}.  (4-6) number of objects selected in this redshift range and stellar mass bin from the CDFS, COSMOS, and UDS \zfourge\ data.  (7) effective radius of prognenitors measured from CANDELS WFC3 F160W imaging. (8) \sersic\ index measured from CANDELS WFC3 F160W imaging.  (9-10) rest-frame $U-V$ and $V-J$ color indices measured from \zfourge\ multiwavelength data.   (11) rest-frame luminosity at 2800~\AA\ derived from the \zfourge\ data.  \ed{Errors on (4--11) are the 68\% percentile range of the distribution. }  (12) fraction of quiescent galaxies, defined as the ratio of the number of galaxies with quiescent $U-V$ and $V-J$ colors to the total number in each bin. \ed{Errors on (12) are derived using a bootstrap Monte Carlo simulation.}}
\ifsubmode
\end{deluxetable}
\else
\end{deluxetable*}
\fi

Figure~\ref{fig:massvol} shows that the \zfourge\ dataset is well
matched to track the stellar-mass evolution of MW and M31--like
galaxies over $0.5 < z < 3$.  At lower redshifts, $z$$<$0.5,  the
\zfourge\ dataset lacks sufficient volume to track the evolution of
galaxies with $M_\ast \gsim 10^{11}$~\msol\ down to $z$=0.
However, the \zfourge\ data is sensitive to the progenitors of these
galaxies to $z \approx 3.3$, where the expected progenitor mass equals
the stellar-mass completeness limit.   Similarly,
figure~\ref{fig:massvol} shows that \zfourge\ is complete for
progenitors of MW-sized galaxies to $z\approx 2$.   Formally, the
stellar-mass completeness limit is derived for red, passive stellar
populations, whereas the mass limit for blue, star-forming galaxies is
lower by about 1 dex.  As we show below, nearly all the MW progenitors
at these redshifts fall in the latter category, so we expect to track
MW progenitors out to $z > 2.5$.  Therefore, within the single,
homogeneous \zfourge\ dataset, we are able to track the evolution of
the MW-mass and M31-mass galaxies over a long baseline in time, which
corresponds to the majority of the galaxies' formation history.

We select $M^\ast$ progenitors from \zfourge\ in bins of comoving
volume and mass as illustrated in figure~\ref{fig:massvol}.
\ed{Table~\ref{table:main} lists the redshift intervals, and the central
  value of the stellar mass used to select the subsamples.
 We select progenitors of the M31- and MW-mass
galaxies that have stellar mass within $\pm$0.25 dex of the central
value of stellar mass in each redshift.}   Our choice of $\pm$0.25 dex in
stellar mass is motivated by both the differences in mass evolution
based on different abundance matching (or constant number density)
methods, and also based on the scatter in the stellar mass of the
progenitors of present-day galaxies \citep[see \eg,][]{behr13c}.
\ed{At higher redshift the interval in redshift of the bins
increases as a compromise between comoving volume and lookback time
spanned by each bin.}   
In the lowest redshift bins there is overlap between the MW and M31
progenitors subsamples (i.e., the boxes overlap in
figure~\ref{fig:massvol}).  This is acceptable because the scatter in
the progenitor mass evolution means that the descendants of the
galaxies in the overlap region may become either MW- or M31-mass
galaxies at $z\sim 0$ \citep[again, see discussion in,
\eg,][]{behr13c}.  
\ed{ Table~\ref{table:main} lists the number of galaxies from
  each \zfourge\ field, and the median mass of the galaxies
  selected in each subsample.}   
\ed{Table~\ref{table:main} also lists the median
and 68 percentile range on the distribution of the $U-V$ and $V-J$
rest-frame color, and the effective radius and \sersic\ index of the
galaxies in each subsample of $M^\ast$ galaxies.}

\section{Color Evolution of M$^\ast$ Galaxies}

\ifsubmode
\begin{figure} 
\epsscale{0.8} 
\else
\begin{figure*} [p]
\epsscale{1.1} 
\fi
%
%\epsscale{0.7} 
%\plotone{colorcolor_mos_nil.ps}

\plotone{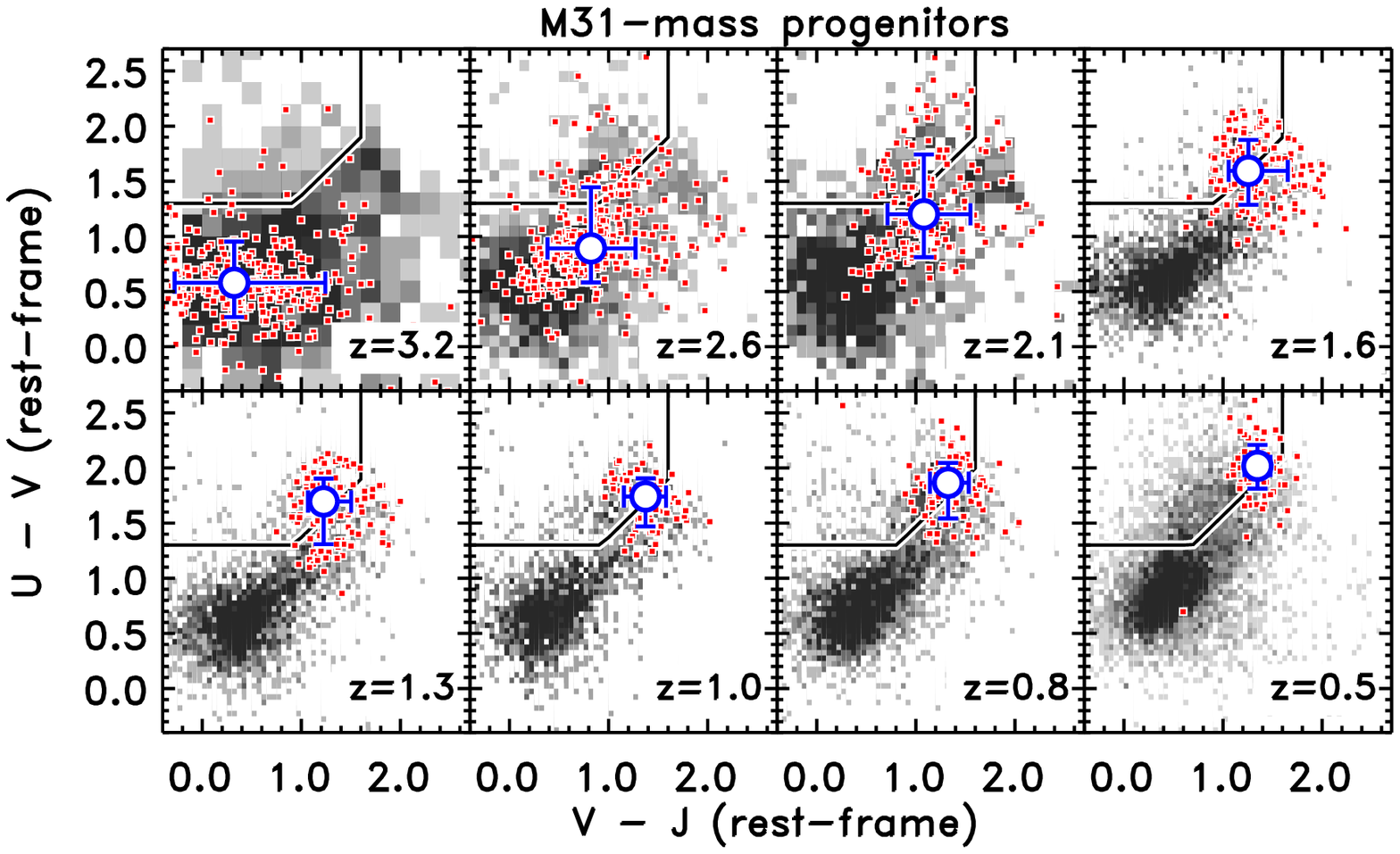}
\plotone{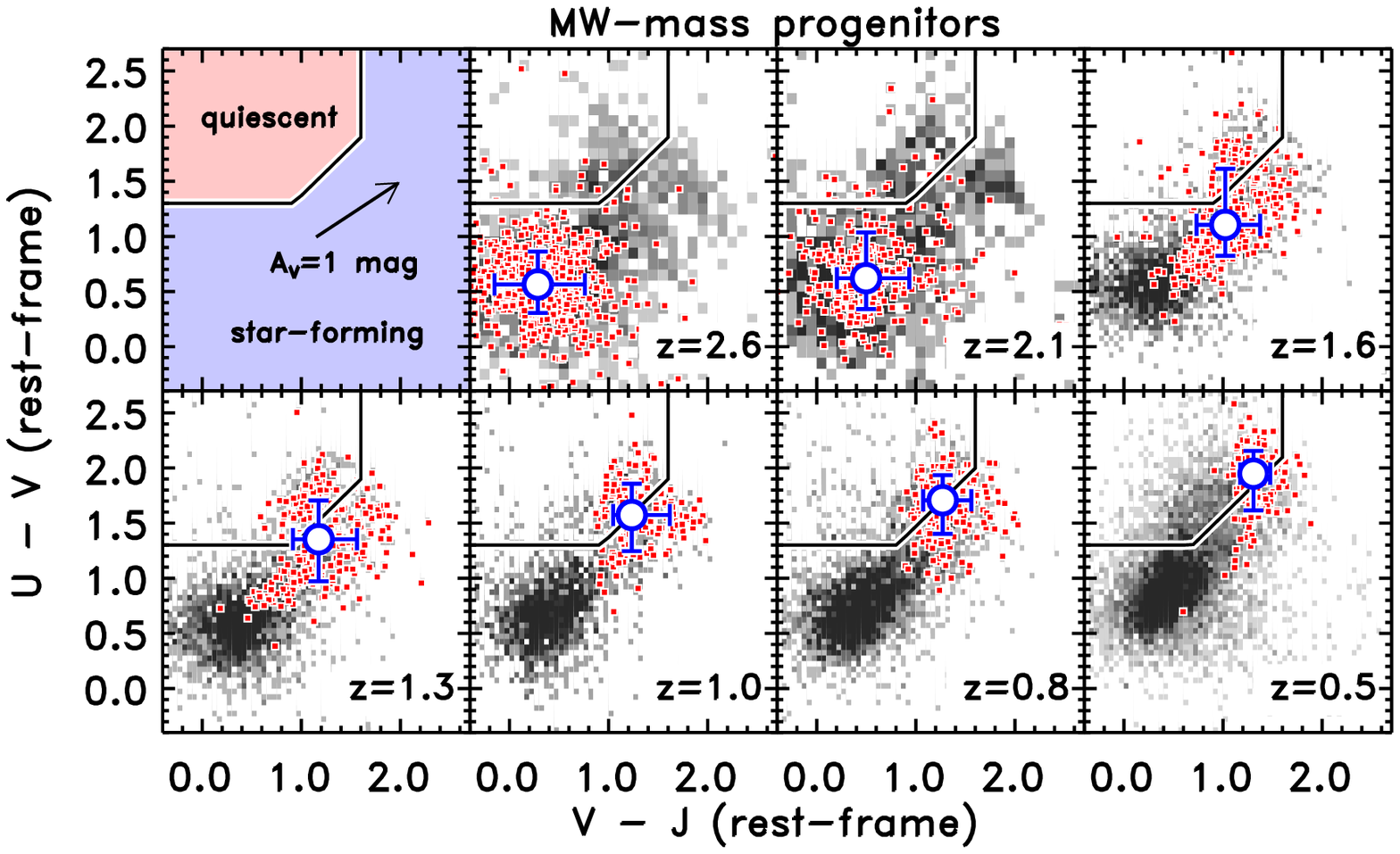}
\caption{The evolution of rest-frame $U-V$ and $V-J$ color
distributions of progenitors of M31-mass and MW-mass galaxies.    The
polygon in each panel delineates ``quiescent'' galaxies (upper left
region) from ``star-forming'' (everywhere else) using the definition
of \citet{will09}, as indicated in the top left panel of the MW-mass
progenitor plot.  The arrow in that panel shows the effects on the
colors for $A_V$=1 mag of dust attenuation for the starburst dust
model \citep{calz00}.   In both plots the grey-scale increases with
the density of \ed{all galaxies in the \zfourge\ catalogs} that have
those rest-frame colors in each redshift bin.    The red points in
each panel show the $M^\ast$ galaxy progenitors.   The top figure
shows the color evolution of M31-mass progenitors, and the bottom
figure shows MW-mass progenitors.   The large circles in each bin show
the median value of the $M^\ast$ galaxy progenitors in each panel.
\ed{The error bars show the 68 percentile range of rest-frame colors
for the M31-mass and MW-mass progenitors.}}\label{fig:colorcolor}  
\epsscale{1} 
\ifsubmode
\end{figure}
\else
\end{figure*}
\fi

Figure~\ref{fig:colorcolor} shows the evolution of the rest-frame
$U-V$ and $V-J$ colors (a $UVJ$ diagram) of the M31- and MW-mass
galaxy progenitors from $z=0.5$ to 3.     The rest-frame $UVJ$ color-color plane
separates galaxies that are actively star-forming from those in
quiescent phases of evolution \citep[e.g.,][]{labbe05,wuyts07,will09,whit11,papo12}.
Galaxies that fall in the star-forming region of the $UVJ$ diagram
have high current SFRs compared to their past
average.  In contrast, galaxies in the quiescent region of the $UVJ$
diagram have current SFRs much lower than
their past average.   The sequence of star-forming
galaxies follows dust attenuation as the colors move along the $UVJ$
diagram from relatively unattenuated galaxies with blue $U-V$ and
$V-J$ colors to those with higher dust attenuation and red $U-V$ and
$V-J$ colors. 

\begin{figure}
\ifsubmode
\else
\epsscale{1.1}
\fi
\plotone{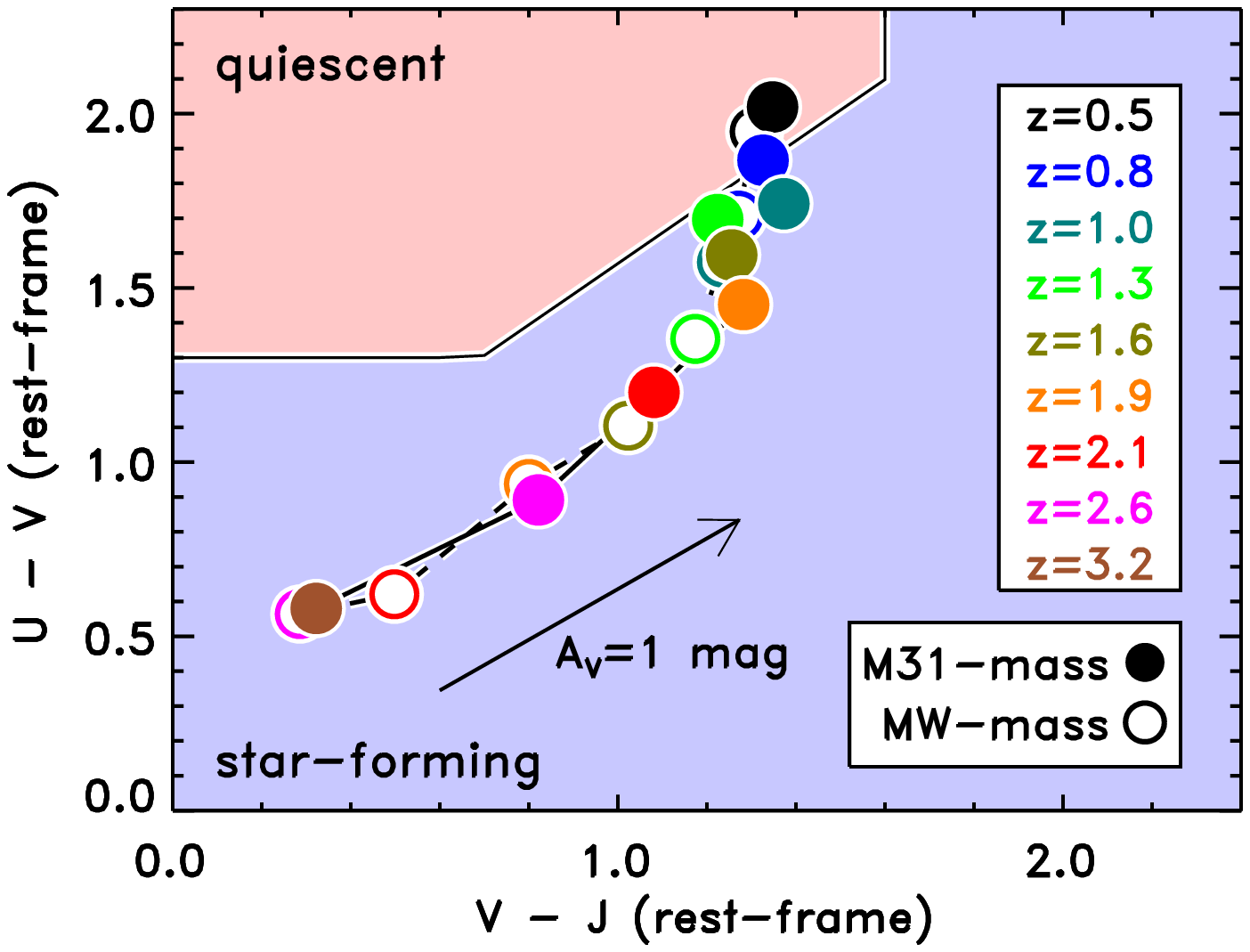}
\caption{The evolution in the median $U-V$ and $V-J$ color for the MW-mass
  progenitors (open symbols connected by dashed lines) and M31-mass
  progenitors (filled symbols connected by solid lines).
  Each point shows the median color at each redshift as indicated by
  color (see figure inset).   As in figure~\ref{fig:colorcolor}, the
  polygon denotes the region in the upper left populated by quiescent
  galaxies.   The arrow shows the effects on the colors for $A_V$=1
  mag of dust attenuation for the starburst dust model
  \citep{calz00}.  The MW- and M31-mass progenitors follow similar color
  evolution, but the more massive M31-mass progenitors evolve earlier (at
  higher redshift) compared to the less massive MW progenitors.}\label{fig:colorz}
\end{figure}

Figure~\ref{fig:colorz} shows that both the M31- and MW-mass progenitors have
similar evolution in their median $U-V$ and $V-J$ color with redshift.
However, the  changes in the galaxies occur at earlier times (higher
redshifts) for the higher-mass M31 progenitors compared to the
lower-mass MW progenitors.    At the highest redshifts ($z \gsim
2.5$), the progenitors are blue in both their $U-V$ and $V-J$ colors,
indicating they are star-forming with relatively low dust attenuation.
As the population moves to lower redshifts ($1.6 \lsim z \lsim 2.5$),
the $U-V$ and $V-J$ colors become redder, indicating they are
star-forming but with higher dust attenuation, and there are
essentially \textit{no} blue, unattenuated galaxies.   At redshifts
less than about $z\lsim 1$, the progenitors become a mix of galaxies
with dust-attenuated star-forming galaxies and quiescent objects whose
star-formation is quenching.   The color evolution reflects this as an
increasing portion of the evolution occurs as a reddening of the
median $U-V$ color.   As a result, by $z\lsim 0.5$ the majority of
both the MW and M31 progenitors have crossed into the quiescent
region, indicating these galaxies have either quenched their star formation,
or are forming stars at rates much less than their past average.

\begin{figure}
\ifsubmode
\else
\epsscale{1.1}
\fi
\plotone{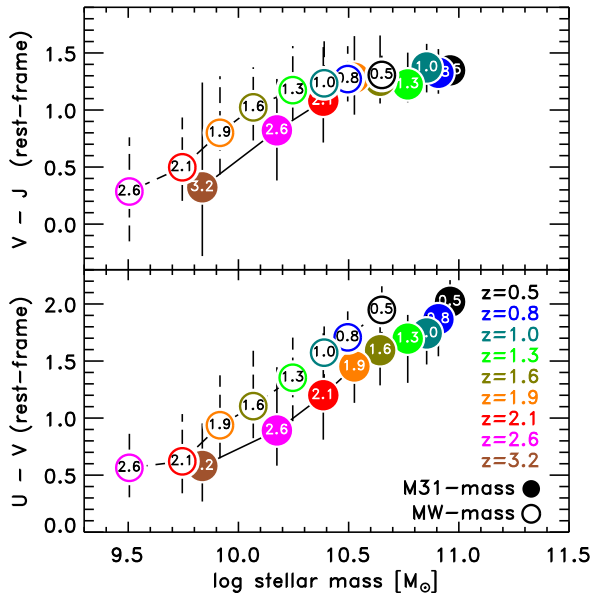}
\caption{The evolution in the median $U-V$ and $V-J$ color as a
  function of stellar mass for the MW-mass
  progenitors (open symbols connected by dashed lines) and M31-mass
  progenitors (filled symbols connected by solid lines).
  Each point shows the median color at each redshift as indicated by
  its label and  color (see figure inset).  The error bars span the
  inter-68 percentile of each subsample.    While the MW- and M31-mass progenitors follow similar color
  evolution, the M31-mass progenitors achieve redder
  colors at higher fixed stellar mass compared to the lower mass
  MW-mass progenitors.}\label{fig:colormassz}
\end{figure}

While the M31-mass and MW-mass progenitors follow similar
color-evolutionary paths, they do so at different stellar masses.
Figure~\ref{fig:colormassz} shows the evolution between the median
rest-frame colors as a function of mass and redshift.    At
fixed \textit{stellar mass} the massive M31-mass progenitors have
bluer rest-frame $U-V$ and $V-J$ colors compared to the less massive
MW-mass progenitors.  Therefore the color-evolution is dependent both
on stellar mass and redshift.  

\ifsubmode
\begin{figure}
\else
\begin{figure}
\epsscale{1.1}
\fi
\plotone{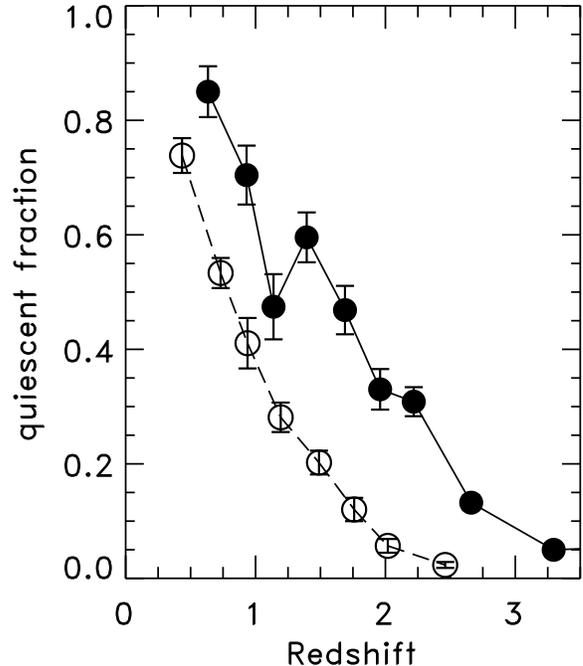}
\caption{The evolution of the fraction of quiescent galaxies for
$M^\ast$ galaxies as a function of redshift.   The filled circles and
solid line show the evolution of the M31-mass progenitors.  The
open circles and dashed line show the evolution of the MW-mass
progenitors.   The
quiescent fraction of the MW and M31 progenitors increases with
decreasing redshift, although at all redshifts the M31
progenitors have a higher quiescent fraction.  }\label{fig:qfrac}
\epsscale{1} \ifsubmode
\end{figure}
\else
\end{figure}
\fi
 
The M31 progenitors become quiescent at earlier times (higher
redshift) compared to the lower-mass MW progenitors.
Figure~\ref{fig:qfrac} shows the evolution in the quiescent fraction
of the M31 and MW-progenitors, where the quiescent fraction is defined
as the ratio of the total number of galaxies  falling in the
``quiescent'' region of the $UVJ$ diagram to the total number of
galaxies in each progenitor sample at each redshift.     \ed{We
derived uncertainties on the quiescent fraction using a bootstrap
monte carlo simulation.  We reconstructed each subsample repeatedly
with the same number of galaxies in each reconstruction, but randomly
drawing from each subsample with replacement.  We took as the
uncertainty the standard deviation using the normalized median
absolute deviation of the distribution of quiescent fractions from the
reconstructions \citep[see,][]{bram08}.  The quiescent fractions are
listed in table~\ref{table:main}.}  At all redshifts, the quiescent
fraction of the M31 progenitors is higher.  For both the M31 and MW
progenitors, as they become quiescent, their stellar populations
homogenize.   This is evident from the low scatter on the quiescent
fraction and the $UVJ$ colors (where the error bars span the inter-68
percentile), where the low scatter implies similar mass-dominated
stellar population ages within each subsample. 

\section{Morphological Evolution of the $M^\ast$ 
Galaxies}

\ifsubmode
\begin{figure}
\epsscale{0.8}
\else
\begin{figure}
\epsscale{1.1}
\fi
\plotone{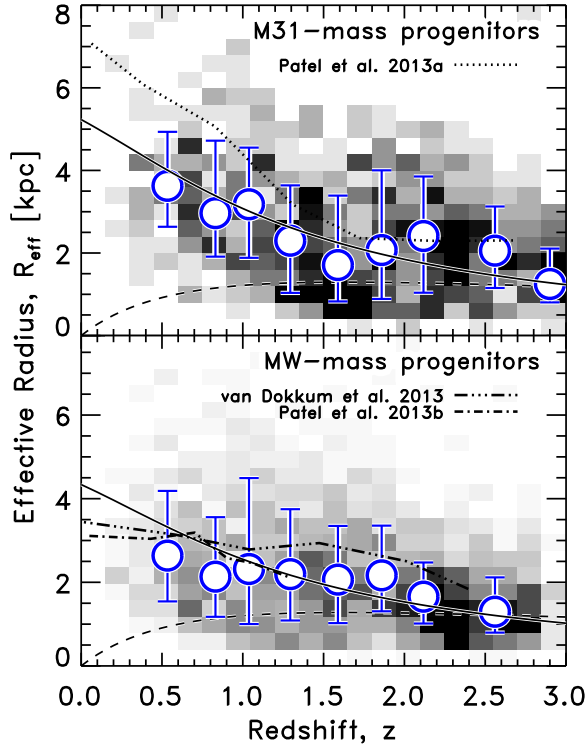}
%\plotone{mwreff_mw.ps}
\caption{Size evolution of M31-mass (top panel) and MW-mass (bottom
panel) galaxies.  In each panel, the shaded regions show the density
of data points that fall in that bin.   The large blue circles are the
median in bins of redshift, and \ed{the error bars show the 68\%-tile
range of the distribution}.  The dashed line shows the FWHM \hst/WFC3
PSF.  The solid line shows a fit to the medians where the effective
size scales with the inverse Hubble parameter, $R_\mathrm{eff} \propto
H(z)^{-1}$.  \ed{The curves show relations for other galaxy samples
taken from the literature.}}\label{fig:reff} \epsscale{1} \ifsubmode
\end{figure}
\else
\end{figure}
\fi

%The CANDELS \hst\ imaging of the \zfourge\ fields provides structural
%information on the galaxies in our M31 and MW progenitor
%samples.   
%

\ifsubmode
\begin{figure}
\epsscale{0.8}
\else
\begin{figure}
\epsscale{1.1}
\fi
\plotone{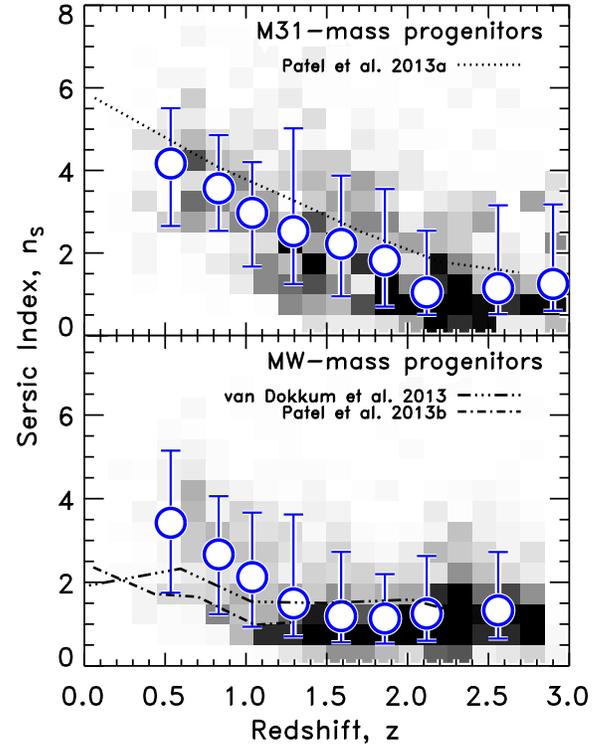}
%\plotone{mwn_mw.ps}
\caption{\sersic\ index evolution of M31-mass (top panel) and
MW-mass (bottom panel) galaxies.   In each panel, the shaded regions
show the density of data points that fall in that bin.  The large blue
circles are the median in bins of redshift, and the error bars show
the \ed{68\%-tile range on the distribution}. \ed{The curves show
  relations for other galaxy samples taken from the literature.}  }\label{fig:n} \epsscale{1} \ifsubmode
\end{figure}
\else
\end{figure}
\fi

Figure~\ref{fig:reff} shows that the size evolution of the M31-mass
and MW-mass progenitor samples.    At $z > 2$ the M31-mass progenitors
are small ($1-2$~kpc) albeit with significant scatter about the
median.  \ed{Although there is significant scatter about the median,
the error on the medians are much smaller (as there are more than 60
galaxies in each bin of redshift).   This yields a measurement of the
rate of size growth in these galaxies that is generally consistent
with the growth of disks within galaxy halos, where $\reff \propto
H(z)^{-1}$ \citep[under the assumption of a constant halo spin
parameter,][]{mo98,ferg04}   Quantitatively, a fit of the function
$\reff \propto H(z)^{-1}$ to the data for the M31-mass progenitors
extrapolates to $\reff(0) = 5.2\pm 0.1$~kpc at $z=0$.  A fit to a more
generic function, $\reff \propto (1+z)^\gamma$ returns $\gamma = -1.0
\pm 0.05$.  Similarly, the MW-mass progenitors are uniformly small at
$z \gsim 2$, with $\reff \sim 1-2$~kpc.   A fit to the data assuming
$\reff \propto H(z)^{-1}$ extrapolates to $\reff(0) = 4.3 \pm
0.05$~kpc at $z=0$.  A fit to the function $\reff \propto
(1+z)^\gamma$ returns $\gamma = -0.9 \pm 0.05$.    There
is clear evidence for size growth in both the M31-mass and MW-mass
progenitors, consistent with the growth of disks.  This is
consistent with other studies \citep[see,][and discussion
below]{patel13a,patel13b,vandokkum13}.}

The median \sersic\ index of the $M^\ast$ galaxy progenitors evolves
smoothly with redshift, as shown in figure~\ref{fig:n}.  At the
highest redshifts, both the M31 and MW progenitors have low \sersic\
indices, with median $\langle \nsersic \rangle \simeq 1$ for $z > 2$
for the M31-mass progenitors, and $z > 1.5$ for the MW-mass
progenitors, consistent with exponentially declining disk-like
surface-brightness profiles.  As the galaxies evolve to lower
redshift, both the M31- and MW-mass progenitors begin a monotonic
increase in their median \sersic\ indices with decreasing redshift,
from a median value of $\langle \nsersic \rangle \simeq 1$ at $z=2$ to
$\langle \nsersic \rangle = 4$ at $z=0.5$ for the M31-mass
progenitors, and $\langle \nsersic \rangle \simeq 1$ at $z=1.3$ to
$\langle \nsersic \rangle =3.5$ at $z=0.5$ for the MW-mass
progenitors.   Equating larger values of $\nsersic$ with bulge
formation, it is during these periods when $M^\ast$ galaxies grow
their spheroids. 

\ifsubmode
\begin{figure}
\else
\begin{figure*}[p]
\epsscale{1.1}
\fi
\begin{center}
\ifsubmode\includegraphics[width=7in,angle=90]{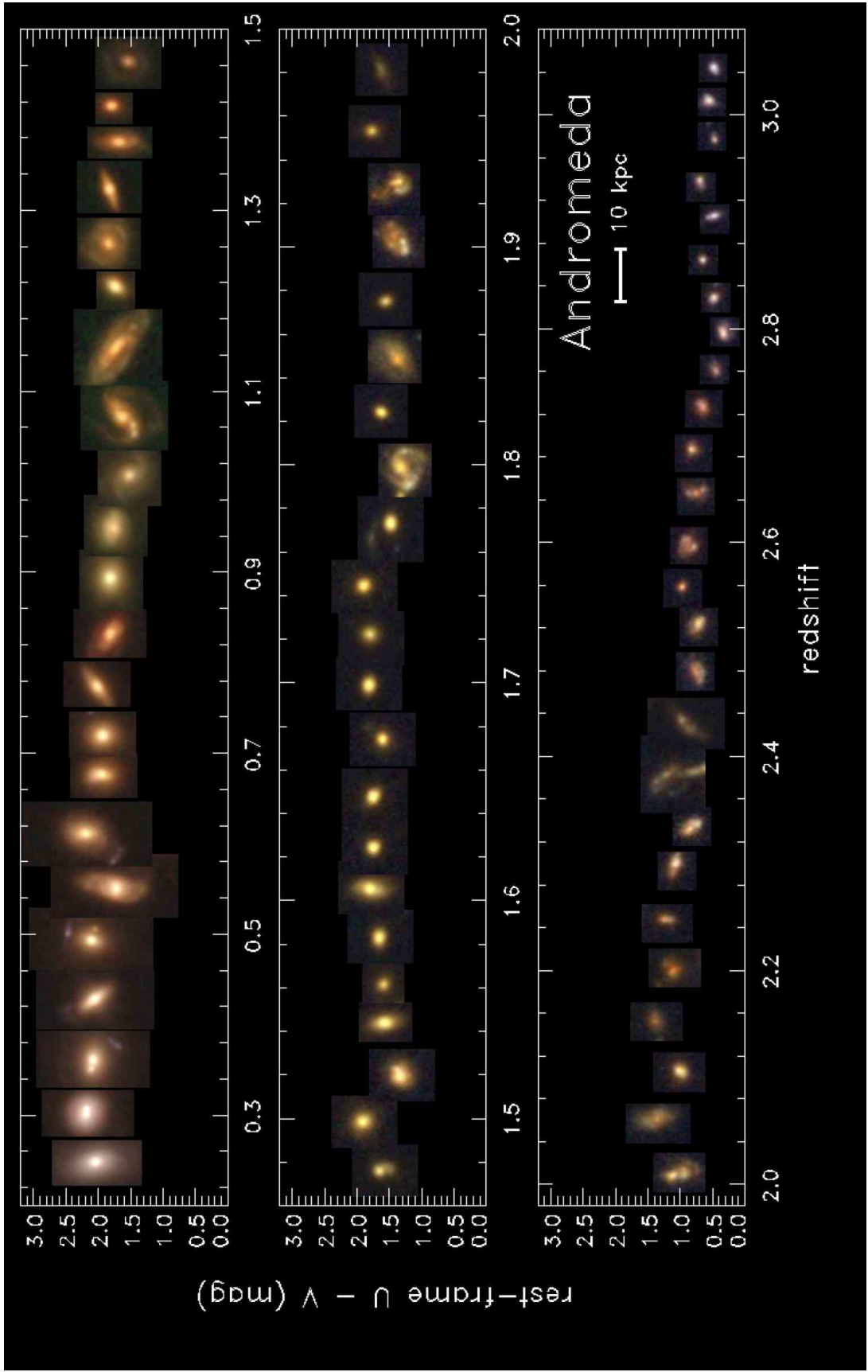}
\else \includegraphics[width=9in,angle=90]{fig12.eps}
\fi
\caption{Examples of progenitors of an M31-mass galaxy from $z=3$ to
the $z=0.5$.   Each galaxy is selected such that it has the
(approximate) median $U-V$ and $V-J$ color derived for all progenitors
in a given redshift bin (see Table~\ref{table:main}).  Each
false-color image shows the approximate rest-frame $U$, $B$, $V$-band
(blue, green, red, respectively) using the ACS (\acsb\acsv\acsi\acsz)
and WFC3 (\wfcj\wfch) band closest to rest-frame $UBV$ at each
redshift \citep[for this reason we show only progenitors from the
CDF-S sample, because this full complement of \hst\ imaging does not
exist for the COSMOS nor UDS \zfourge\ fields,
see][]{grogin11,koek11}.  The images are placed at their measured
rest-frame $U-V$ color and redshift (slight adjustments in redshift
are made for presentation purposes, but the rank order of the galaxies
is unchanged).  The image sizes are scaled to the same fixed physical scale
where the inset shows a scale of 10 kpc.}\label{fig:imgAndro}
\end{center}
\epsscale{1}
\ifsubmode
\end{figure}
\else
\end{figure*}
\fi

\ifsubmode
\begin{figure}
\epsscale{0.75}
\else
\begin{figure*}[p]
\epsscale{1.1}
\fi
\begin{center}
\ifsubmode \includegraphics[width=7in,angle=90]{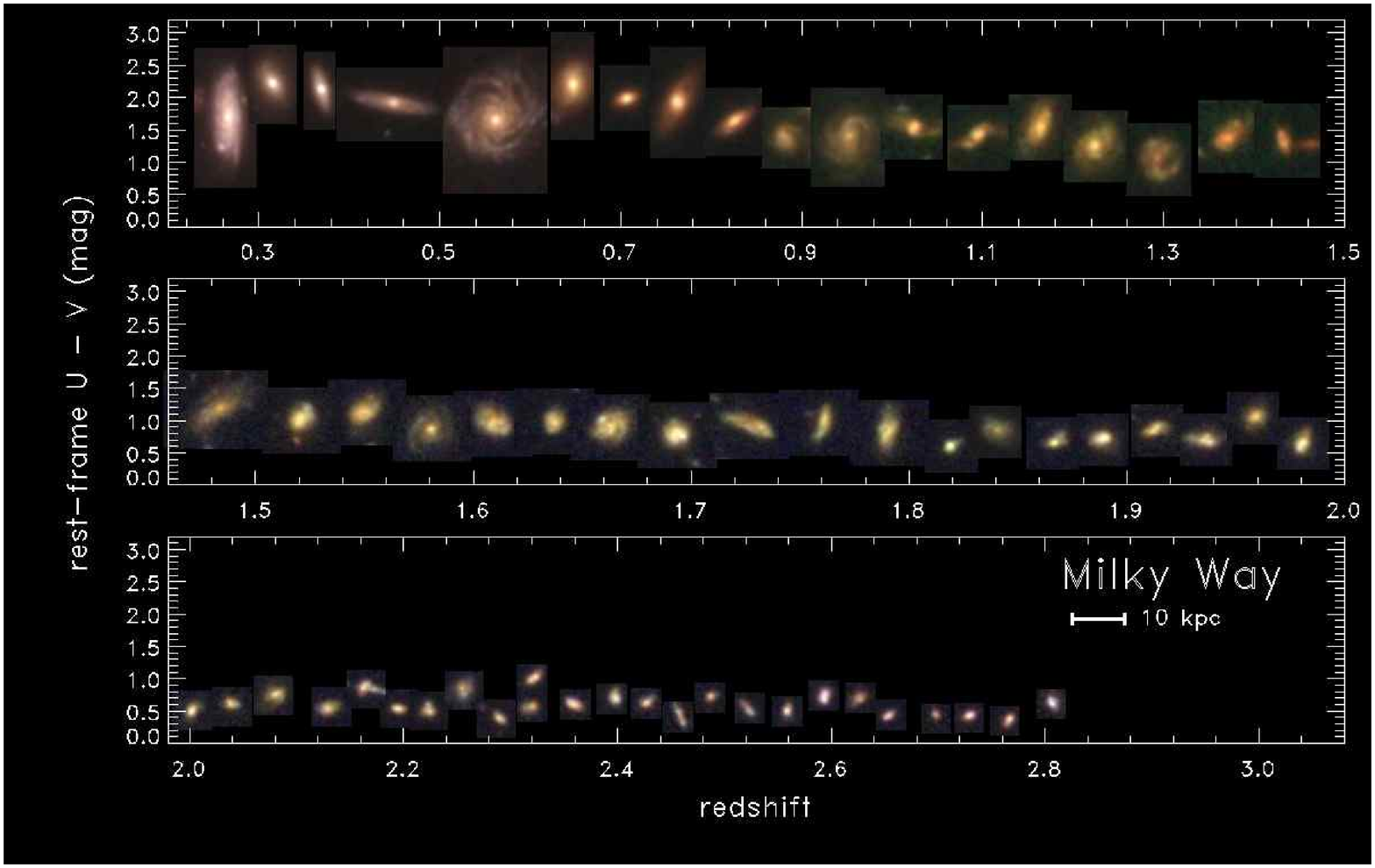}
\else  \includegraphics[width=9in,angle=90]{fig13.eps}
\fi
\caption{Same as figure~\ref{fig:imgAndro}, for progenitors of
  MW-mass galaxies from $z=3$ to the $z=0.5$.  
}\label{fig:imgMW}
\end{center}
\epsscale{1}
\ifsubmode
\end{figure}
\else
\end{figure*}
\fi
 
As with the size evolution, the evolution in the \sersic\ index
is not driven by the morphological $K$-corrections arising from
using a fixed observed band (WFC3 F160W).   The
\sersic\ indices derived from the WFC3 F105W change only slightly from
the values from the WFC3 F160W imaging, and this change does not
affect the qualitative conclusions.     

\ed{Our results compare favorably with those in the literature.}
\citet{patel13a} considered the size evolution of galaxies selected to
have constant number density, $\log (n/\mathrm{Mpc}^{-3}) = -3.9$,
slightly less common (and therefore more massive) than the
M31-mass progenitors in our sample (the latter have number density
closer to $\log [n/\mathrm{Mpc}^{-3}] = -3.4$, see
figure~\ref{fig:ndense}).  \ed{Figures~\ref{fig:reff} and \ref{fig:n}
show that the galaxies in the Patel et al.\ sample have larger
effective radii and larger \sersic\ indexes compared to our values for
the M31-mass progenitors at the same redshift .   This is likely a result of the higher
stellar masses of the galaxies studied by Patel et al.\ compared to
the M31-mass sample here. }  The differences could possibly be related to
band-shifting effects (``morphological $K$-corrections''), as Patel et
al.\ use ACS F814W imaging for their sample at $z < 1$ where we use
WFC3 F160W \citep[see discussion in ][]{vanderwel14}.  However, there
is only a slight increase in the effective radii of our sample derived
with the WFC3 F105W compared to the F160W band (an increase of the
median of at most $\approx$10\% at $z=0.5$).  Therefore, the choice of
bandpass for the effective sizes does not change the qualitative
conclusions. 

\ed{\citet{vandokkum13} considered the evolution of the \sersic\ index
of MW-sized progenitors, and derive somewhat different evolution.  As
illustrated in Figures~\ref{fig:reff} and \ref{fig:n} shows that by
$z\sim 2$ the galaxies in their sample have already achieved higher
\sersic\ index, $\nsersic\sim 2$, higher than those in our sample,
with somewhat higher effective radii at higher redshifts.   It is
likely this is because the van Dokkum et al.\ sample is selected to
have constant number density, which yields higher stellar mass
progenitors at $z\sim 2$  ($\log M_\ast / M_\odot \gsim 10$).  In
comparison, the progenitors in our MW-mass sample are lower in stellar
mass by about a factor of 2 at $z\sim 2$. Because the evolution is
such that at fixed redshift the median \sersic\ index increases with
stellar mass, the difference in stellar mass between the samples
likely accounts for the difference in \sersic-index evolution.}

\ed{\citet{patel13b} traced the structural evolution of star-forming
MW-mass progenitors from $z\sim 0$ to $z = 1.5$.  Their sample was
selected using the inferred growth from the star-forming sequence
\citep[see][]{leit12}.  The galaxies in their progenitor sample have
lower stellar mass (by about 0.2 dex) compared to our MW-progenitor
sample.  Figures~\ref{fig:reff} and \ref{fig:n} show while the
evolution in effective radius is similar between their sample and
ours,  the galaxies in their sample have weaker evolution in \sersic\
index, with $\langle \nsersic \rangle \simeq 1.5-2$ at $z\sim 0.4$
compared to our finding of $\langle \nsersic \rangle =3.4$ at $z=0.5$.
This difference is likely due entirely to the fact that the Patel et
al.\ samples are star-forming only, with lower stellar mass. }

The visual morphology of the progenitors of M31-mass and MW-mass
galaxies encapsulates their evolution, as illustrated in  in
figures~\ref{fig:imgAndro} and \ref{fig:imgMW}.  Each figure shows
(approximate) rest-frame $UBV$ images of galaxies from the progenitors
subsamples that have  the median $U-V$ and $V-J$ colors derived in
figure~\ref{fig:colorcolor} for the M31- and the MW-mass subsamples,
respectively.  Specifically, we select from the progenitors those
galaxies satisfying $(\Delta_{UV}^2 + \Delta_{VJ}^2)^{1/2} \leq 0.3$
mag, where $\Delta_{UV} \equiv \langle(U-V)\rangle - (U-V)$ and
$\Delta_{VJ} \equiv  \langle (V-J) \rangle - (V-J)$ is the color
difference between the median color and the color of each galaxy the
MW- and M31-mass progenitors in each redshift bin.    \ed{We selected
galaxies at random from the subsample that satisfies this color
selection.  We visually inspected cases where galaxies overlapped on
the figure, rejecting objects to ensure that the figure captures the
diversity of morphology.}  At the highest redshifts, the M31- and
MW-mass progenitors are small and blue, with visual morphologies
similar to UV-selected LBGs \citep[\eg,][]{papo05}.  Starting around
$z\sim 2$ the progenitors become more diffuse and nebulous, sometimes
with redder cores and blue outskirts, and some showing star-forming
(blue) clumps \citep[similar in morphology to those in other
star-forming galaxies at these epochs, \eg,][]{elme08}.  Starting
around $z\sim 1-1.5$ more mature morphological structures form, and
some of the progenitors develop spiral arms, and spheroid/disk
combinations: it appears that $z\sim 1$ is the epoch where $M^\ast$
galaxies begin to populate the ``Hubble sequence''.\footnote{Some of
the M31-mass progenitors at $z\sim 1.5-1.7$ show compact (spheroidal)
morphologies.  These may be outliers as these images are all taken
from the GOODS-S, which is known to host an overdense large-scale
structure at $z\sim 1.6$ \citep{kurk09,giav11}.  There is evidence
that galaxies in overdense regions have accelerated morphological
evolution, \citep[\eg,][]{papo12,bass13,delaye14}, which may account
for the more early-type morphologies of the M31 progenitors at this
redshift in this field.}\label{footnote}   At $z \lsim 1$ the
morphologies of the $M^\ast$ progenitors have matured, and  all the
progenitors show either early-type morphologies and/or bulge-dominated
disks, including examples of ``grand design spirals''.     

Comparing figures~\ref{fig:imgAndro} and \ref{fig:imgMW}, with those of
``modern-day'' $M^\ast$ galaxies from SDSS at $0.02 < z < 0.03$ in
figures~\ref{fig:sdssAndro} and \ref{fig:sdssMW} above, we see they
dovetail nicely.   The morphologies of the $z\sim 0.5$ $M^\ast$
galaxies show examples of both spheroid galaxies, bulge-dominated disk
galaxies, and grand design spirals. 

\section{Evolution of the IR emission and the SFR}

\subsection{Stacked IR Images and the Total IR Emission}\label{section:ir}

\ifsubmode
\begin{figure}
\else
\begin{figure*}
\epsscale{1.1}
\fi
\plotone{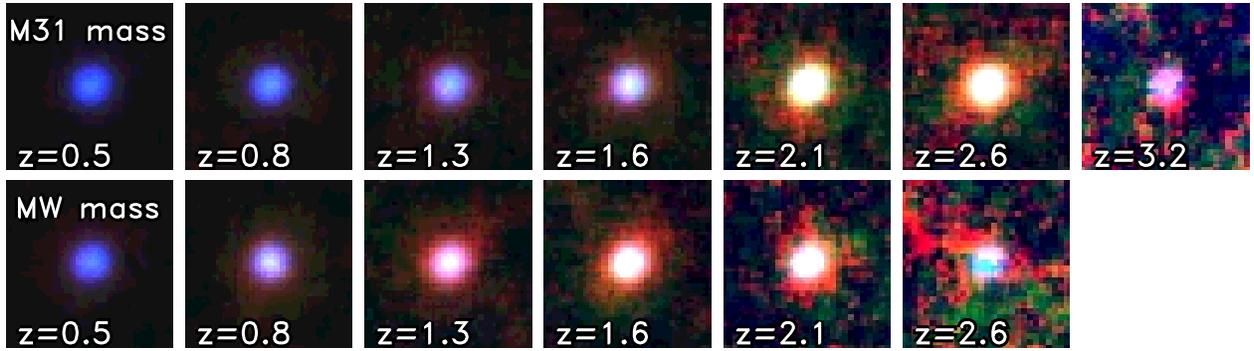}
\caption{Stacked false-color infrared images for the progenitors in the
  M31-mass and MW-mass galaxy subsamples.    The top
 row shows the redshift evolution of M31-mass progenitors and
 the bottom row shows the redshift evolution of MW-mass progenitors.
 In each panel, the false-colors show the 
stacked IRAC 3.6 + 4.5 \micron\ flux density (blue),  the stacked MIPS
24~\micron\ flux density (green), and the mean of the PACS 100 and 160
 \micron\ flux density (red).  The scaling in the false-color images is
tilted so that a source will appear white if it has a spectral energy
distribution that follows $f_\nu \sim \lambda^{1.5}$.  The images have not been
 matched in their point-response-functions (PRFs).  The PRFs between
 MIPS 24~\micron\ and Herschel 100~\micron\ are about equal (both have
 PRF FWHM  $\simeq$6\arcsec) , but are  larger than that of IRAC (PRF
 FWHM $\simeq$1.6\arcsec).  }\label{fig:IRrgb}
\epsscale{1}
\ifsubmode
\end{figure}
\else
\end{figure*}
\fi

The majority of the galaxies in both the M31 and MW progenitor samples
are undetected in the \spitzer/MIPS 24~\micron\ and \herschel/PACS 100
and 160~\micron\  imaging.   The fraction of sources detected at
24~\micron\ ($>5\sigma$) is less than 50\% for $z < 1$, and this
declines to $<$30\% at $z=1.9$ and $<$10\% at $z > 2$.  At all
redshifts, the objects \textit{detected} in the MIPS and PACS data
represent only the most luminous galaxies at each redshift, forming a
biased subset.  Therefore, to study the evolution in the far-IR
emission from the galaxies in the M31- and MW-progenitor samples we
stack the mid- and far-IR data to improve the effective depth.  By
doing this we lose the ability to study galaxies on an
object-by-object basis, but we gain the ability to measure the average
24, 100, and 160~\micron\ emission for these samples to fainter flux
densities than otherwise possible.    Stacking techniques have proven
valuable to study the IR emission from faint galaxy populations
\citep[\eg,][]{dole06,zheng06,zheng07,dye07,huynh07,papo07,nlee10,rodi10,schr14}. \citet{papo07}
showed that for sources with flux densities about a factor of two
lower than the formal S/N=5 detection limit, one can recover the
average stacked fluxes accurate to better than 10\% for sufficiently
large samples ($N > 100-200$ objects).  

Here we used the method described in \citet{papo07} to stack the IR
emission for all the galaxies in each redshift bin for the M31- and
MW-mass progenitor samples.   We stacked the galaxies in the MIPS
24~\micron, and PACS 100 and 160~\micron\ data using the $M^\ast$
progenitor samples in each redshift bin discussed in
\S~\ref{section:samples}.  We stacked the IRAC data in addition to the
MIPS/PACS data to provide a reference between the stacked far-IR data
and the near-IR data (see below).  We first take a small subimage of
100\arcsec\ $\times$ 100\arcsec\ from the 3.6, 4.5, 24, 100, and
160~\micron\ data for each source to be stacked.  We used a
two-dimensional bilinear interpolation to center the subimage on the
astrometric coordinates of each source.  We then subtracted a local
background from each source using the sky value measured in a
bandpass-dependent annulus using values from the literature
\citep{popesso12,magn13}$^{\hbox{\scriptsize\ref{footnote:irac},\ref{footnote:mips},\ref{footnote:pacs}}}$
centered on each source.  Following \citet{papo07} we rotated each
subimage randomly by 0$^\circ$, 90$^\circ$, 180$^\circ$, or
270$^\circ$ to suppress any image artifacts, which tend to be aligned
in either detector rows or columns.    We stack the images, taking the
mean of all pixels (after clipping $>$3$\sigma$ outliers) that
contribute to a given pixel in the final stack.    We measured the
flux densities in circular apertures \ed{of radius $2\farcs4$,
$2\farcs4$,  $3\farcs5$, 7\arcsec, and 12\arcsec,  correcting for
light outside those apertures by multiplying the fluxes by factors of
1.20, 1.22,  2.57, 1.47, and 1.45 for IRAC 3.6, 4.5~\micron, MIPS 24
\micron, and Herschel 100, 160~\micron,
respectively}, using values from documentation \footnote{http://irsa.ipac.caltech.edu/data/SPITZER/docs/irac\label{footnote:irac}}$^{\hbox{\scriptsize,}}$\footnote{http://irsa.ipac.caltech.edu/data/SPITZER/docs/mips\label{footnote:mips}}$^{\hbox{\scriptsize,}}$\footnote{http://herschel.esac.esa.int/twiki/pub/Public/PacsCalibrationWeb/pacs\_bolo\_fluxcal\_report\_v1.pdf\label{footnote:pacs}}
and the literature \citep{popesso12,magn13}.  Table~\ref{table:lir}
lists the measured flux densities from the stacks.

We estimated uncertainties for the stacked flux densities using a
bootstrap Monte Carlo simulation.  For each subsample, we repeated the
stacking procedure 100 times.  Each time we constructed a new
subsample equal in size to the original subsample, but which contained
randomly selected galaxies from the original subsample with
replacement.   We then recomputed the flux density from the stack of
the random subsample.  The estimate of the uncertainty is the standard
deviation of the distribution of the measured flux densities from
these Monte Carlo simulations (in this way the estimated uncertainties
are the uncertainty on the median).  Table ~\ref{table:lir} lists
these uncertainties.

We also stacked on random positions in the images as a secondary
estimate of the uncertainties on the flux densities.    \ed{For our
stacking we did not remove flux from detected IR sources because some
of the $M^\ast$ progenitors are directly detected, and we did not want
introduce biases by erroneously subtracting the light from sources in
our sample that may be undetected but otherwise blended with nearby
sources.  Rather we tested how this affects our measurements by using
the random stack to measure any bias resulting from stacking
procedue.}    We stacked a number of random pointings equal to the
number of sources in each subsample (\ed{where we have made no
requirement on the location of the random pointings, which can fall on
or near detected IR sources}), and we repeated this step $>$500 times.
We then computed the mean and standard deviation of the flux density
distribution from the random stacks.  \ed{For large numbers of
sources, $N > 10^5$, the mean flux density from the stack at random
positions is small, $\langle f_\nu \rangle$ = ($-0.015$, $-0.015$,
$0.6$, $-24$, 47 $\mu$Jy) at (3.6,  4.5, 24, 100, and 160 \micron),
respectively.  In all cases this bias is much lower than the
uncertainty, with S/N $<$ 0.3 for the size subsamples here.}
Similarly, for all samples except those at the highest redshift, the
standard deviation from the random stacks is much smaller than the
uncertainties computed above.  Therefore, variations of the IR
emission within the progenitor samples dominate the uncertainties of
the stacked flux densities.  In the cases of the highest redshift
subsamples ($z=3.1$ for the M31-mass progenitors and $z=2.5$ for the
MW-mass progenitors), the uncertainty from the random stacks at (24,
100, and 160~\micron) are approximately equal to those from our Monte
Carlo simulation above.  This implies that the image noise (a
combination of sky and confusion noise) dominates the uncertainties in
the stack for these subsamples.   Because our results from the Monte
Carlo simulation on the galaxy subsamples include both the variation
in the IR emission of subsample and effects from the sky noise, we
adopt these uncertainties here.  

%\ed{While the stackes on random positions are sensitive to biases in
%  the data and stacking procedure, they do not account for biases
%  resulting 

\ifsubmode
 \begin{deluxetable}{lccccccccc}
\tabletypesize{\scriptsize}
\else
 \begin{deluxetable*}{lccccccccc}
\fi
\tablecaption{Stacked IR Properties, SFRs, and implied gas fractions of $M^\ast$-Galaxy Progenitors\label{table:lir}}
\tablecolumns{10}
\tablewidth{0pc}
\tablehead{
\colhead{} &
\colhead{} &
\colhead{$F_\nu(3.6\micron)$} &
\colhead{$F_\nu(4.5\micron)$} &
\colhead{$F_\nu(24\micron)$} &
\colhead{$F_\nu(100\micron)$} &
\colhead{$F_\nu(160\micron)$} &
\colhead{\lir} & 
\colhead{SFR} & 
\colhead{} \\
\colhead{z} & 
\colhead{N} & 
\colhead{($\mu$Jy)} & 
\colhead{($\mu$Jy)} & 
\colhead{($\mu$Jy)} & 
\colhead{($\mu$Jy)} & 
\colhead{($\mu$Jy)} & 
\colhead{($10^{11}\ \lsol$)} & 
\colhead{(\msol yr$^{-1}$)} & 
\colhead{$f_\mathrm{gas}$} \\
\colhead{(1)} & 
\colhead{(2)} & 
\colhead{(3)} & 
\colhead{(4)} & 
\colhead{(5)} & 
\colhead{(6)} & 
\colhead{(7)} & 
\colhead{(8)} & 
\colhead{(9)} & 
\colhead{(10)} }
\startdata
\multicolumn{10}{c}{M31-mass Progenitors} \\[3pt]
0.5 & 100 &  115.6 $\pm$    5.1\phn\phn &  \phn  81.4 $\pm$    4.5\phn\phn &   50.6 $\pm$   10.7\phn\phn &         1217\phd\phn $\pm$          247 &         2410\phd\phn $\pm$  \phn          612 & 0.22 $\pm$ 0.03 &  2.8 $\pm$  0.8 & 0.03 $\pm$ 0.01 \\
0.8 & 115 &  \phn  70.6 $\pm$    3.6\phn\phn &  \phn  48.1 $\pm$    2.4\phn\phn &   73.9 $\pm$   16.3\phn\phn &         1240\phd\phn $\pm$          276 &         2973\phd\phn $\pm$  \phn          828 & 0.77 $\pm$ 0.11 &  8.4 $\pm$  1.7 & 0.06 $\pm$ 0.02 \\
1.0 & \phn 78 &  \phn  57.5 $\pm$    2.4\phn\phn &  \phn  46.4 $\pm$    2.5\phn\phn &  129.9 $\pm$   19.3\phn\phn &         2188\phd\phn $\pm$          403 &         5918\phd\phn $\pm$         1118 & 2.22 $\pm$ 0.22 & 23.6 $\pm$  3.6 & 0.13 $\pm$ 0.03 \\
1.3 & 136 &  \phn  32.0 $\pm$    1.6\phn\phn &  \phn  29.7 $\pm$    1.3\phn\phn &   60.1 $\pm$  \phn    7.4\phn\phn &         1124\phd\phn $\pm$          283 &         2524\phd\phn $\pm$  \phn          706 & 2.53 $\pm$ 0.26 & 26.7 $\pm$  4.1 & 0.14 $\pm$ 0.04 \\
1.6 & 175 &  \phn  16.7 $\pm$   0.76\phn &  \phn  18.4 $\pm$   0.71\phn &   55.0 $\pm$  \phn    7.8\phn\phn &  \phn          844\phd\phn $\pm$          127 &         2171\phd\phn $\pm$  \phn          403 & 3.30 $\pm$ 0.30 & 34.6 $\pm$  4.7 & 0.19 $\pm$ 0.04 \\
1.8 & 209 &  \phn \phn    9.6 $\pm$   0.35\phn &  \phn  11.3 $\pm$   0.37\phn &   76.7 $\pm$  \phn    8.0\phn\phn &  \phn          894\phd\phn $\pm$          111 &         2409\phd\phn $\pm$  \phn          366 & 4.74 $\pm$ 0.34 & 48.9 $\pm$  4.9 & 0.30 $\pm$ 0.06 \\
2.1 & 175 &  \phn \phn    5.5 $\pm$   0.22\phn &  \phn \phn    6.4 $\pm$   0.30\phn &   60.0 $\pm$  \phn    3.6\phn\phn &  \phn          528\phd\phn $\pm$  \phn   96.3 &         1284\phd\phn $\pm$  \phn          302 & 3.89 $\pm$ 0.22 & 41.3 $\pm$  4.6 & 0.36 $\pm$ 0.08 \\
2.6 & 350 &  \phn \phn    3.2 $\pm$   0.11\phn &  \phn \phn    3.7 $\pm$   0.13\phn &   28.5 $\pm$  \phn    2.6\phn\phn &  \phn          409\phd\phn $\pm$  \phn   60.8 &  \phn          763\phd\phn $\pm$  \phn          245 & 4.78 $\pm$ 0.37 & 52.2 $\pm$  8.1 & 0.50 $\pm$ 0.13 \\
3.2 & 406 &  \phn \phn    1.1 $\pm$   0.06 &  \phn \phn    1.3 $\pm$   0.06 &  \phn    2.8 $\pm$  \phn    1.0\phn\phn &  \phn  \phn  76.0 $\pm$  \phn   43.9 &  \phn  \phn  45.3 $\pm$  \phn          189 & 1.48 $\pm$ 0.48 & 21.8 $\pm$ 11.8 & 0.47 $\pm$ 0.30 \\\hline
\\ \multicolumn{10}{c}{MW-mass Progenitors} \\[3pt]
0.5 & 191 &  \phn  74.9 $\pm$    2.5\phn\phn &  \phn  51.8 $\pm$    2.2\phn\phn &   61.7 $\pm$   14.1\phn\phn &         1374\phd\phn $\pm$          201 &         3129\phd\phn $\pm$  \phn          597 & 0.28 $\pm$ 0.03 &  3.3 $\pm$  0.7 & 0.05 $\pm$ 0.02 \\
0.8 & 221 &  \phn  38.1 $\pm$    1.3\phn\phn &  \phn  26.5 $\pm$   0.90\phn &   89.6 $\pm$   12.6\phn\phn &         1562\phd\phn $\pm$          196 &         3463\phd\phn $\pm$  \phn          467 & 0.96 $\pm$ 0.07 & 10.1 $\pm$  1.2 & 0.13 $\pm$ 0.03 \\
1.0 & 165 &  \phn  27.0 $\pm$   0.88\phn &  \phn  21.4 $\pm$   0.95\phn &  103.5 $\pm$  \phn    9.4\phn\phn &         1841\phd\phn $\pm$          169 &         4937\phd\phn $\pm$  \phn          528 & 1.81 $\pm$ 0.10 & 18.8 $\pm$  1.8 & 0.23 $\pm$ 0.04 \\
1.3 & 290 &  \phn  13.1 $\pm$   0.51\phn &  \phn  12.5 $\pm$   0.46\phn &   47.1 $\pm$  \phn    1.9\phn\phn &  \phn          922\phd\phn $\pm$  \phn   82.5 &         2367\phd\phn $\pm$  \phn          302 & 2.03 $\pm$ 0.07 & 21.4 $\pm$  1.8 & 0.32 $\pm$ 0.06 \\
1.6 & 356 &  \phn \phn    6.5 $\pm$   0.19\phn &  \phn \phn    7.4 $\pm$   0.21\phn &   45.3 $\pm$  \phn    2.7\phn\phn &  \phn          668\phd\phn $\pm$  \phn   76.3 &         1404\phd\phn $\pm$  \phn          278 & 2.69 $\pm$ 0.14 & 28.7 $\pm$  3.2 & 0.46 $\pm$ 0.09 \\
1.8 & 291 &  \phn \phn    4.2 $\pm$   0.18\phn &  \phn \phn    4.6 $\pm$   0.22\phn &   38.5 $\pm$  \phn    3.2\phn\phn &  \phn          485\phd\phn $\pm$  \phn   58.3 &         1395\phd\phn $\pm$  \phn          305 & 2.53 $\pm$ 0.17 & 28.3 $\pm$  4.6 & 0.55 $\pm$ 0.15 \\
2.1 & 370 &  \phn \phn    2.2 $\pm$   0.12\phn &  \phn \phn    2.3 $\pm$   0.12\phn &   16.4 $\pm$  \phn    1.8\phn\phn &  \phn          238\phd\phn $\pm$  \phn   58.6 &  \phn          526\phd\phn $\pm$  \phn          189 & 1.50 $\pm$ 0.15 & 18.9 $\pm$  5.3 & 0.54 $\pm$ 0.22 \\
2.6 & 869 &  \phn \phn    1.1 $\pm$   0.04 &  \phn \phn    1.1 $\pm$   0.04 &  \phn    5.2 $\pm$  \phn    1.0\phn &  \phn  \phn  67.7 $\pm$  \phn   36.4 &  \phn          202\phd\phn $\pm$  \phn          141 & 0.85 $\pm$ 0.15 & 12.8 $\pm$  5.8 & 0.57 $\pm$ 0.32 \\\hline
\enddata
\tablecomments{(1) Redshift of bin, (2) number of objects stacked in
each bin, (3-7) measured flux density from the IRAC 3.6\micron,
4.5\micron, MIPS 24\micron, PACS 100\micron, and 160\micron,
respectively, (8) total IR luminosity derived from the stacked MIPS
and PAC photometry, (9) total SFR derived from the \lir\ and
rest-frame luminosity at 2800\AA, (10) implied gas-mass fraction,
defined as $M_\mathrm{gas} / (M_\ast + M_\mathrm{gas})$.}
\ifsubmode
\end{deluxetable}
\else
\end{deluxetable*}
\fi

Figure~\ref{fig:IRrgb} shows images of the stacked IRAC,
MIPS, and PACS data for each of the redshift subsamples for the M31-
and MW-progenitor samples.  While
$K$-corrections in the bandpasses persist, in general the blue color
of the progenitors at $z\sim 0.5$ means that direct starlight
traced in the IRAC bands exceeds the reprocessed
dust emission traced by the far-IR data.  The white color of the
progenitors at higher redshifts, $2.1 \leq z \leq 2.6$ for M31, and
$1.6 \leq z \leq 2.1$ for the MW, means that the dust-reprocessed
emission contributes more to the bolometric emission than the direct
starlight. 
 
\ifsubmode
\begin{figure}
\epsscale{0.67}
\else
\begin{figure*}
\epsscale{0.8}
\fi
%
%\plottwo{fig15a.ps}{fig15b.ps}
\plotone{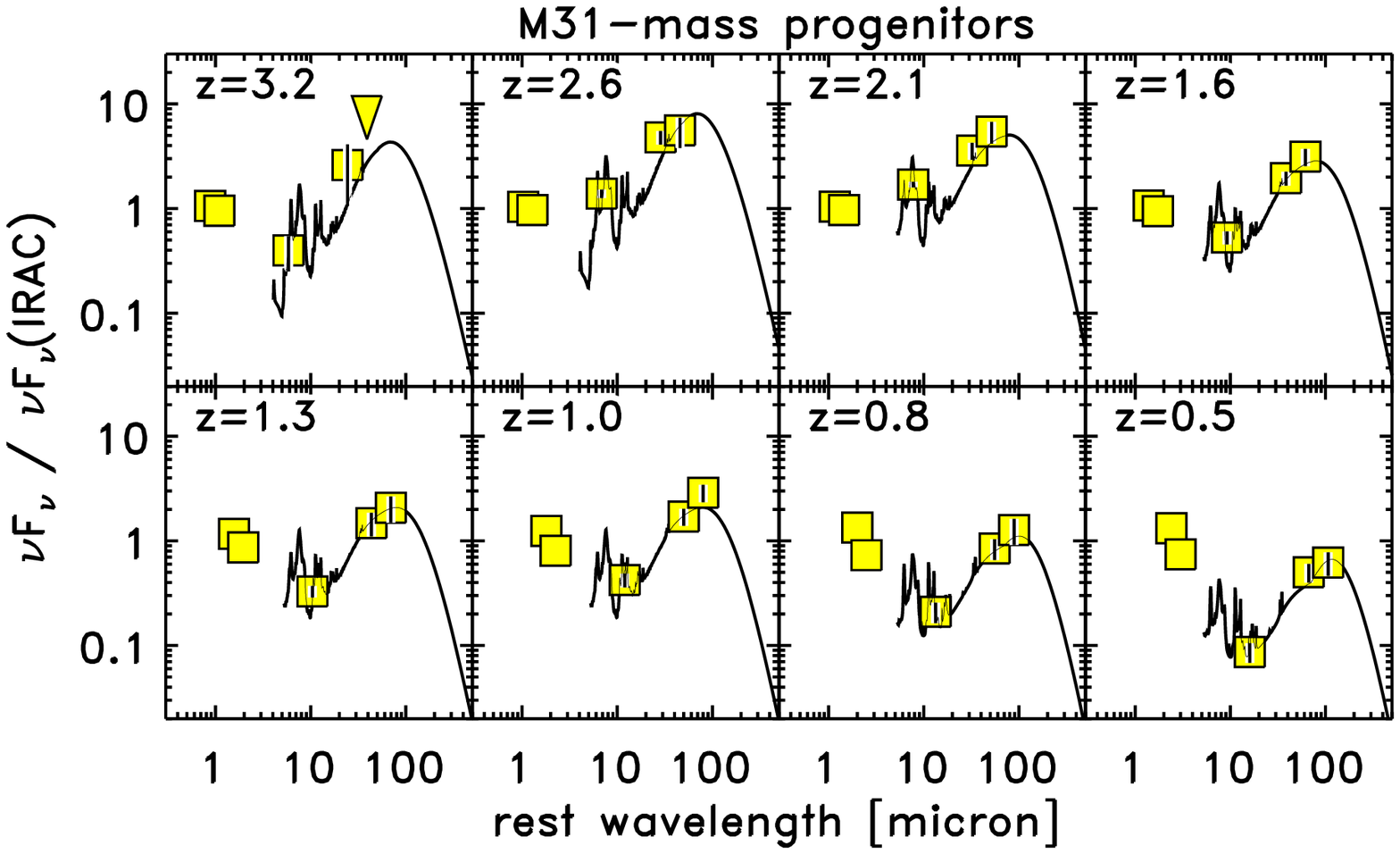}
\plotone{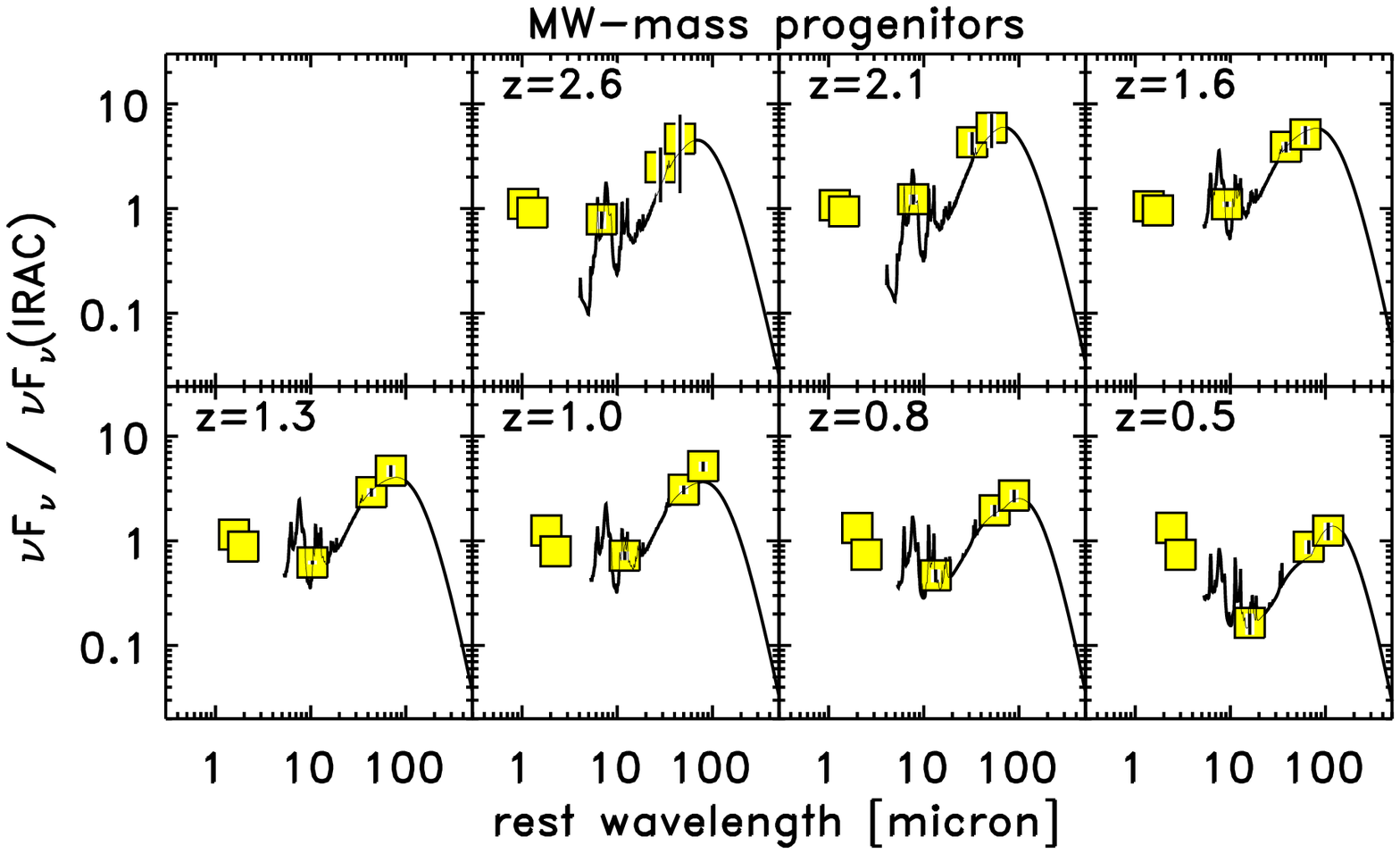}
\caption{The evolution of the IR spectral energy distributions of
$M^\ast$ galaxy progenitors.  The top figure shows the evolution of
M31-mass progenitors, and the bottom figure shows the evolution of the
MW-mass progenitors.  In each panel of each figure, the yellow-filled
boxes show the stacked flux densities from IRAC 3.6, 4.5~\micron, MIPS
24~\micron, and PACS 100, 160~\micron\ for each subsample in redshift,
normalized to the average IRAC 3.6 and 4.5~\micron\ data.  The
uncertainties are derived using the Monte Carlo simulation described
in the text.  Downward triangles show 2$\sigma$ upper limits for IR
flux densities with measured $f_\nu/\sigma(f_\nu)< 1$.    The line in
each panel shows the best-fit \citet{rieke09} IR template spectral
energy distribution.  We infer the total $\lir = L(8-1000\micron)$ IR
luminosity by integrating the best-fit template for each
subsample.}\label{fig:lirSed} \epsscale{1} \ifsubmode
\end{figure}
\else
\end{figure*}
\fi
 
For each subsample of the M31- and MW-mass progenitors, we fitted
template IR spectral energy distributions \citep{rieke09} to the
stacked 24, 100, and 160~\micron\ flux densities and their
uncertainties from the stacks to estimate total IR luminosities, $\lir
= L(8-1000\micron)$, and their uncertainties (where these are the
uncertainties on the median).   These are listed in
Table~\ref{table:lir}.  Because the IR flux densities cover the mid-IR
(the Wein side of the thermal emission) to the far-IR wavelengths
(covering the peak of the thermal dust emission), the choice of IR
spectral templates makes only a small difference in the total IR
luminosities. Using the \citet{chary01} or \citet{dale05} templates
changes the derived IR luminosities by $<$30\% (0.1 dex).  Our choice
to use the Rieke et al.\ templates is motivated by the fact that these
templates better reproduce the far-IR flux ratios of observed galaxies
at high redshifts \citep[see Rieke et al.\
and][]{ship13}. 

Figure~\ref{fig:lirSed} shows the evolution of the IR spectral energy
distributions for the M31- and MW-mass progenitors.  At
$z=0.5$ the emission from direct starlight is larger than the
dust-reprocessed emission for the M31-mass progenitors by a large factor,
and for the MW progenitors the stellar light and dust-reprocessed
emission are comparable.  As the redshift increases, the far-IR
emission increases relative to the near-IR emission, such that the
far-IR emission dominates the bolometric output for $1 \lsim z \lsim
2.6$.  At the highest redshifts, $z=3.2$ for M31 and
$z=2.6$ for the MW, the contribution from the
thermal far-IR emission to the bolometric emission declines.  Because these galaxies
are all star-forming with very blue rest-frame $UVJ$ colors, there
is less dust (and lower obscuration) in these galaxies.

\ifsubmode
\begin{figure}
\epsscale{0.8}
\else
\begin{figure}
\epsscale{1.1}
\fi
\plotone{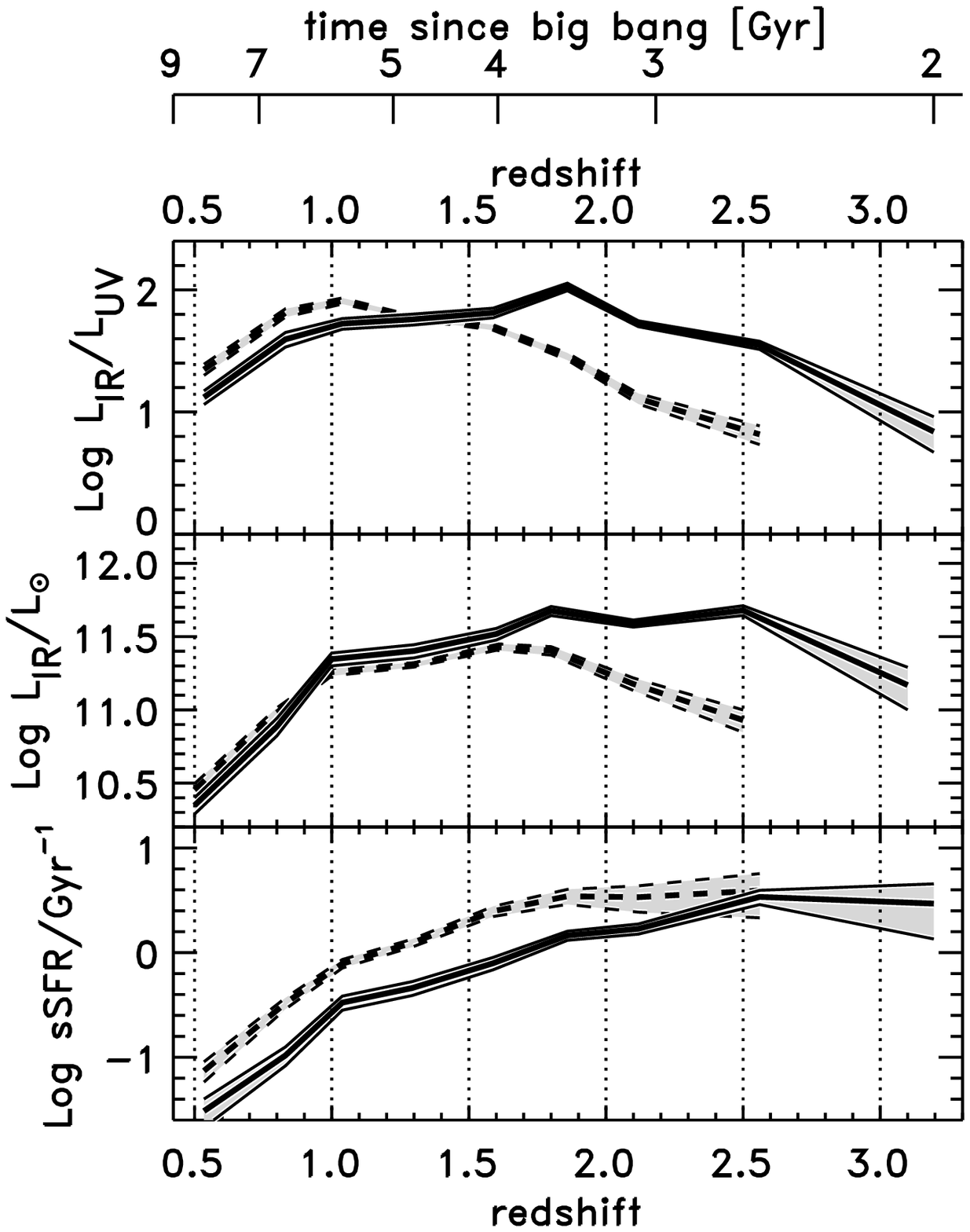}
\caption{The evolution of the IR/UV ratio (top panel) total IR
luminosity (middle panel), and specific SFR (sSFR; bottom panel).
In all panels, the solid lines show the evolution of the M31-mass
progenitors, the dashed lines show the evolution of the MW-mass
progenitors, and the estimated uncertainties on the median value.
}\label{fig:lirevol}
\epsscale{1} 
\ifsubmode
\end{figure}
\else
\end{figure}
\fi

Figure~\ref{fig:lirevol} shows the evolution of the inferred \lir\
from $z$=0.5 to 3.  The IR luminosity of both the M31- and MW-mass
progenitors rise from the highest redshifts, $z = 2.5$ (3.2 for the
M31-mass progenitors)
to reach a plateau for $1 < z \lsim 2$ (2.5 for the case of M31-mass progenitors).
Both progenitor populations show a marked decrease in \lir\ from $z=1$
to 0.5.   The M31-mass progenitors have a higher peak \lir\ (by more
than a factor of two) compared to that of the MW progenitors.  The M31
progenitors reach their peak earlier in their evolution, at $z\simeq
1.8-2.5$,  and sustain this peak for a longer duration, compared to
the MW progenitors, which reach their peak IR luminosity at $z\simeq
1.6$.

Although AGN are likely rare in the $M^\ast$ progenitor samples, they
do occur in hosts with stellar masses $\gsim$$10^{10}$~\msol\
\citep[\eg,][]{kauf04,card10}, and could contribute to the stacked IR
emission.  We matched all objects in the progenitor samples to the
CANDELS-matched X-ray catalogs for the CDF-S and COSMOS fields
\citep[and in prep]{koce12}.  Likely AGN, with $L_X$$>$$10^{43}$ erg s$^{-1}$, account for $<$8\% of the progenitors in any
subsample at any redshift.  We excluded these sources and restack the
IR luminosity for each subsample, which lowered the implied IR
luminosities by $<$0.1 dex for all samples at all redshifts.
Because star-formation likely contributes to the IR emission even in
galaxies hosting AGN \citep[\eg,][]{ship13}, on average AGN do not
strongly contribute to the IR luminosity for the MW-mass and M31-mass
progenitors.

\subsection{Evolution of the IR/UV Ratio}

The total IR luminosities allow us to study the
evolution of the IR/UV luminosity ratio, defined here as the ratio of
the total IR luminosity to the rest-frame luminosity at 2800~\AA\
(where $L_{2800}$ has no correction for dust attenuation),  $\lir/L_{2800} \equiv
\lir/L_\mathrm{UV}$.    The top panel of Figure~\ref{fig:lirevol} shows the evolution
of this ratio for the M31- and MW-mass progenitors. 

The evolution in $\lir/L_\mathrm{UV}$ is similar for both the M31- and
the MW-mass progenitors.   At the highest redshifts, $z=3.2$ for the
M31-mass, and $z=2.5$ for the MW-mass progenitors $\lir/L_\mathrm{UV}
< 10$, which is typical for  LBGs \citep{reddy10}.  The
$\lir/L_\mathrm{UV}$ values increase with time (decreasing redshift),
peaking at $\lir/L_\mathrm{UV} \approx 100$ at $z\sim 2$ and $z\sim 1$
for the M31-mass and the MW-mass progenitors, respectively.    These
ratios are more typical of ultra luminous IR galaxies
\citep[ULIRGS,][]{papo06a,reddy10}.   The implication is that these
progenitor galaxies are producing and retaining greater amounts of
dust, which then absorb radiation from star-formation, re-emitting in
it the far-IR.

At lower redshifts, $\lir/L_\mathrm{UV}$ drops to $\approx$10 at $z =
0.5$.  The fact $\lir/L_\mathrm{UV}$ declines implies the IR
luminosity is declining faster than the rest-frame UV.  The SFR is
declining, and/or that there is a reduction in the density of dust in
the galaxy \citep[perhaps as a result of the declining gas density,
see \S~\ref{section:gasFraction}, and a constant
gas-mass--to--dust-mass ratio for fixed metallicity,
e.g.,][]{bell03a}.

\subsection{Evolution of the SFR and specific SFR}

Nearly all the bolometric emission from star-formation is emitted in
the UV and IR \citep[see discussion in, \eg,][]{bell03a}.  We use the
SFR conversion from the combination of the UV and IR luminosities \citep{bell05,papo06a} to estimate the
instantaneous SFRs for the M31- and MW-mass progenitors, 
\begin{equation}
\Psi/\msol\ \mathrm{yr}^{-1} = 10^{-10} (\lir +
3.3L_{2800}) / \lsol, 
\end{equation}
based on the calibration presented by \citet{kenn98}, and the constant
of proportionality is adjusted for the Chabrier IMF assumed
here. Table~\ref{table:lir} gives the derived SFRs.  

Figure~\ref{fig:bc} shows the SFR history for the MW- and M31-mass
galaxies.  The SFRs of the MW- and M31-mass progenitors are already
high ($>10$~\msol\ yr$^{-1}$) at the highest redshifts to which we can
observe them.  The SFRs peak around $z=2-2.5$ for the M31-mass
progenitors (with a peak value of $\Psi \simeq$ 50 \msol\ yr$^{-1}$)
and around $z=1.5$ for the MW progenitors (with a peak value $\Psi
\simeq$ 30 \msol\ yr$^{-1}$).    The SFRs are nearly equal for both
the M31 and MW progenitors at $z\sim 1$, and they decline at about the
same rate to values of a few solar masses per year at $z=0.5$
\citep[and this decline continues to the present; the current SFR of
the MW and M31 galaxies proper is $0.5-1.5$ \msol\ yr$^{-1}$,
see][]{mutch11}.    The observed SFR evolution in figure~\ref{fig:bc}
matches qualitatively with that derived through a complex abundance
matching by \citet[see their figure 6]{behr13a}.  Because these
results are independent and based on very different analyses and
datasets, the level of agreement and the fact that we are settling on
a mean SFR history for galaxies with the stellar masses of M31 and the MW is encouraging
\ed{\citep[see also,][]{patel13b,vandokkum13}.}

Figure~\ref{fig:lirevol} shows the evolution of the specific SFR.  The behavior of the
specific SFR \ed{is} approximately the same for both the M31- and MW-mass progenitors, but they are
offset in redshift.   The M31-mass progenitors have a plateau in $\log$
(sSFR/Gyr$^{-1}) \approx 0.5$ for $z\geq 2.5$, with a steady decline
toward lower redshift. The magnitude of this plateau in sSFR is 
consistent with other studies of star-forming galaxies at $z > 2$
\citep[see][and references therein]{reddy12}.     The MW-mass progenitors
have a similar plateau with the same value for $z \geq 1.8$, and also
show a steady decline toward lower redshift.  This follows from the
fact that at $z \lsim 1$ the SFR evolution is nearly identical for
both the MW- and M31-mass progenitors, but because the MW-mass
progenitors have lower stellar mass, they have higher specific SFRs.

\ifsubmode
\begin{figure}
\epsscale{0.8}
\else
\begin{figure}
\epsscale{1.1}
\fi
\plotone{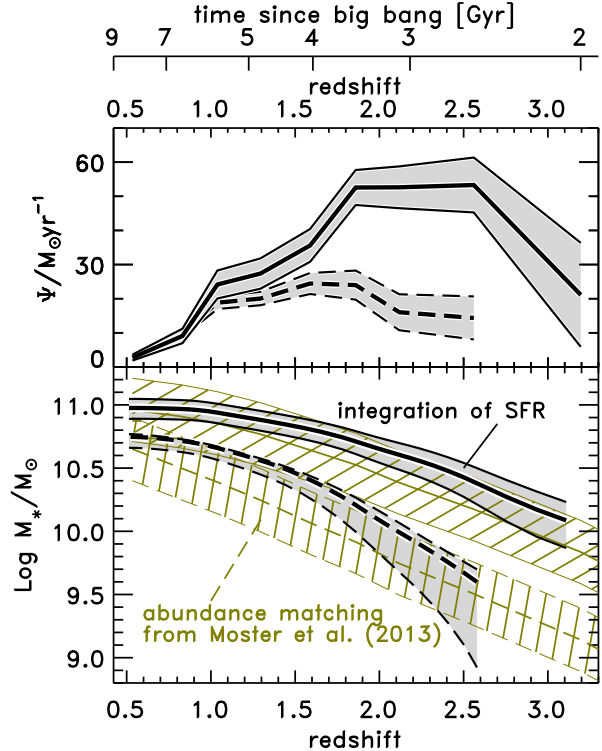}
\caption{The evolution of the SFR, $\Psi$, and the stellar mass
derived from the integrated SFR history.  The top panel shows the SFR
evolution for the M31-mass (solid lines) and MW-mass (dashed lines)
progenitors, where the width of each region corresponds to the
inter-68\% percentile.    The bottom panel shows the integrated SFR
history, derived by using the the SFR evolution with the
\citet{bruz03} stellar population synthesis models.  The shaded
regions correspond to the M31-mass (solid lines) and MW-mass (solid
lines) progenitors,  respectively, as in the top panel.  The hashed
regions in the bottom panel show the stellar-mass evolution from the
abundance matching \citep{moster13} used to the select the progenitor
galaxy samples.  While there is some offset, there is general
agreement between the models and the data.  }\label{fig:bc}
\epsscale{1} \ifsubmode
\end{figure}
\else
\end{figure}
\fi
 
Figure~\ref{fig:bc} shows that the SFR history inferred from the IR
and UV data for the M31- and MW-mass progenitors agrees with the
stellar mass evolution derived by \citet{moster13}, which was used to
select the progenitors themselves.  We obtain the stellar-mass
evolutions by integrating the SFR histories with the \citet{bruz03}
stellar population synthesis model.    At intermediate redshifts, the
integrated SFR histories rise faster than that predicted by the Moster
et al.\ abundance matching.  This may be because our measurements of
the SFR histories have a coarse sampling in cosmic time (averaged over
bins of redshift spanning at least $\approx 10$\% of a Hubble Time at
each redshift), whereas the abundance matching is more continuous with
redshift:  our integrated SFR histories may lack the time resolution
needed to recover the exact stellar-mass evolutions.    Regardless,
the offset is not large (within the $0.25$ dex spread on the stellar
mass history), and the consistency between the stellar mass from the
abundance matching and the integrated SFR history is reassuring that
the SFR history is a reasonable representation of that of $M^\ast$
galaxies such as M31 and the MW. 

\subsection{Evolution of the Implied Gas Fraction}\label{section:gasFraction}

\ifsubmode
\begin{figure}
\else
\begin{figure}
\epsscale{1.1}
\fi
\plotone{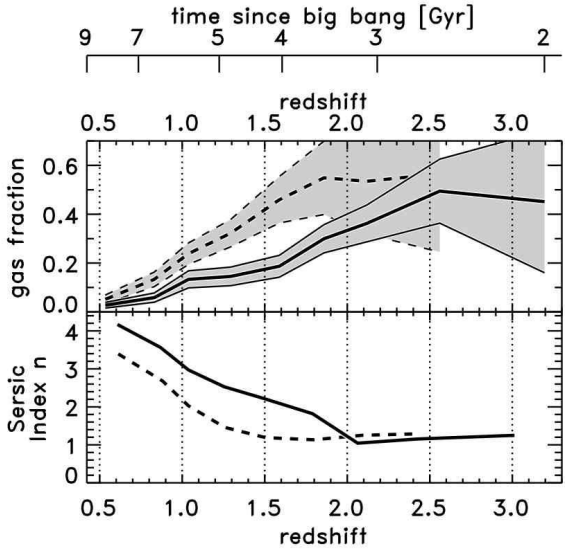}
\caption{Evolution of the implied gas fraction of $M^\ast$ galaxies.
The top panel shows the evolution of the gas fraction, defined as $f_\mathrm{gas} = M_\mathrm{gas} /
(M_\mathrm{gas} + M_\ast)$.  The solid lines show the median and
68 percentile range for the M31-mass progenitors and the dashed lines show
the median and 68 percentile range for the MW-mass progenitors. The bottom
panel reproduces the average evolution of the \sersic\ evolution for
the M31- and MW-mass progenitors as solid and dashed lines,
respectively.   The evolution of the gas fraction is anticorrelated
with the \sersic\ index. }\label{fig:gasevol} \epsscale{1}
\ifsubmode
\end{figure}
\else
\end{figure}
\fi

The surface density of SFR is correlated with the local surface
density of cold gas in galaxies through the established
Kennicutt--Schmidt law \citep{schmidt59,kenn98b}.  Therefore, the
measured SFRs and sizes of the $M^\ast$ galaxies constrain the
gas-mass surface density (and therefore the implied gas mass).
Inverting the relation between the local gas-surface density and the
SFR surface density gives
\begin{equation}\label{eqn:gasmass}
%
%\frac{M_\mathrm{gas}}{5.9 \times 10^{8}\ \msol} = \left(
%
\frac{M_\mathrm{gas}}{6.8 \times 10^{8}\ \msol} = \left(
 \frac{\Psi}{1\ \msol\ \mathrm{yr}^{-1}} \right)^{5/7} \left(
  \frac{r_\mathrm{eff}}{1\ \mathrm{kpc}} \right)^{4/7},
\end{equation}
as in \ed{\citet[see also Conselice et al.\ 2013]{papo11}}.  Using
this equation, we derived gas masses for our progenitor samples using
the effective sizes ($\reff$) and SFRs ($\Psi$) from
tables~\ref{table:main} and \ref{table:lir}.   The gas fraction is
given by comparing the gas masses with the stellar masses by 
\begin{equation}
f_\mathrm{gas} = \frac{M_\mathrm{gas}}{M_\mathrm{gas} + M_\ast}. 
\end{equation}
Table~\ref{table:lir} gives the gas-mass fractions for each of the
progenitor subsamples.

Figure~\ref{fig:gasevol} shows the evolution of these (implied) gas
fractions for the M31- and MW-mass progenitor samples.   The gas
fractions are $f_\mathrm{gas} \simeq 0.4 - 0.6$ at $z > 2.5$ for the
M31-mass progenitors (and at $z > 2$ for the MW-mass progenitors).
Such high gas fractions are consistent with values derived for other
star-forming galaxies at these redshifts, $z > 2$, which have
$f_\mathrm{gas} \sim 0.3-0.5$ for galaxies with stellar masses $\log
M_\ast / \msol = 9.5-10.4$
\citep[\eg,][]{erb06b,daddi09b,forster09,tacc10,tacc13}.

Figure~\ref{fig:gasevol} also shows that the average gas-mass
fractions for the $M^\ast$-galaxy progenitors decline monotonically
with decreasing redshifts.  \ed{There are strong trends in stellar
  mass.}  For the more massive M31-mass galaxies,
this decrease in the gas-mass fraction begins by $z$$\simeq$2.5, whereas
it occurs later (by $z$$\sim$2) for the less massive MW-mass galaxies.
\ed{The direction of this trend is also consisent with \citet{cons13},
  who find gas fractions 0--0.2 for most galaxies with $\log M_\ast/\msol >
  11$ at $z$$>$1.5. }
The decline in the implied gas-mass fractions \ed{of the $M^\ast$
  galaxies} continues with
decreasing redshift.  By $z$$\sim$0.5 the values for the $M^\ast$
progenitor samples are very low:  the median values are
$f_\mathrm{gas}$=0.03 for M31 and 0.05 for the MW.  Furthermore,
figure~\ref{fig:gasevol} shows that the decline in the average
gas-mass fraction is simultaneous with an increase in the \sersic\ index
for the $M^\ast$ progenitor samples.   

Our implied gas-mass fractions for the $M^\ast$ progenitors are
$\langle f_\mathrm{gas} \rangle = 0.03-0.05$ by $z$$\sim$0.5.
\citet{sain11} find that nearby galaxies ($D_L < 200$ Mpc) with
detected H$_2$ molecular gas and stellar masses $\log M_\ast/\msol =
10.7-11$ have implied total cold gas fractions $f_\mathrm{gas} \simeq
0.07-0.11$ (albeit with appreciable galaxy-to-galaxy scatter).
Similar results are found by \citet{morg06}, but see also \citet{young09}.  The observed cold gas fractions from
Saintonge et al.\  are \textit{higher} by about a factor of two
compared to the gas fractions we infer for the $M^\ast$ progenitors
here.  One reason to expect this difference is that the
$M^\ast$-galaxy progenitors may contain additional cold gas, but that
the star-formation efficiency is necessarily low such that this gas
does not contribute to the SFR (as we derive the implied gas fractions
inverting the Kennicutt-Schmidt law).   We discuss this further below.

\section{Discussion}

\subsection{The Growth of $M^\ast$ Galaxies}

\subsubsection{The LBG phase}

To the highest redshifts at which we are able to observe them,  the
MW- and M31-mass galaxy progenitors exist as UV-luminous, star-forming
galaxies with relatively low obscuration at $z\sim 2-2.5$ and $z=3.2$,
respectively.    Their stellar masses, (blue) rest-frame colors, and
small effective sizes are typical of the well-studied ($R \lsim
25$~mag) LBG and Lyman-$\alpha$ Emitter (LAE) populations at these
redshifts
\citep[\eg,][]{stei99,giav02,papo01,papo05,gawi07,shap01,shap11,nils11,varg14}.
\ed{Although at our highest redshift bins, our samples could be biased
against progenitors that are redden by dust or quiescent stellar
populations, we argue this is not the case for the reasons in \S~3.2.
During these star-forming stages more than 50\% of the present-day
stellar-mass is formed in $M^\ast$-mass progenitors. }  This conclusion
is consistent with that of \citet[see also Patel et al.\
2013b]{vandokkum13}.  These redshifts mark the ``LBG phase'' of
MW-mass and M31-mass progenitors, and it seems likely that the main
progenitors of both the MW and M31 would have existed as LBGs
$10-11$~Gyr ago.

During this phase the typical main progenitors of $M^\ast$ galaxies
are star-forming disks ($\nsersic \simeq 1$, fig.~\ref{fig:n}).   At
these redshifts, the implied gas fractions in these disks are high
because the SFRs are high and the scale radii small (\S~7.3 and figure
18).  These disks are likely gas rich and highly turbulent,
similar to dispersion-dominated star-forming disks measured in higher
mass galaxies at these redshifts  \citep[\eg,][]{genzel08,forster11}.
Because the \sersic\ indices show no evidence for evolution at these
redshifts, the progenitors are predominantly disks.  Therefore, it is
either the case that any physical processes capable of transferring
(e.g., clump migration, disk instabilities) or redistributing mass
(e.g., mergers) to a bulge/spheroid are not acting in a substantial
way for $M^\ast$ progenitor galaxies, or that any process that
transfers material to the center must be counterbalanced by continued
mass growth (i.e., via accretion) in the outskirts in such a way that
the \sersic\ index stays low while the total stellar mass and radius
both grow.  

\subsubsection{The Luminous IR--Galaxy Phase}

All indications based on the rest-frame color evolution
(figures~\ref{fig:colorcolor} and \ref{fig:colorz}), the IR-luminosity
evolution, and $\lir/L_\mathrm{UV}$ evolution (figure~\ref{fig:lirevol}) show
that the dust obscuration increases as the $M^\ast$
progenitors form their stars. The measured $\lir$, SFRs, and $\lir /
L_{UV}$ ratios all peak during this period ($1 < z < 2$) and then
they decline at lower redshift ($z \lsim 1$).    At the end of these
phases the $M^\ast$ galaxies have formed $>$75\% of their present-day
stellar mass.    Most of the stars in galaxies such as M31 and the MW
formed their stars during the first $\sim$6 Gyr of the history of the
Universe.  

Figure~\ref{fig:bc} also shows that essentially \textit{all} of the
stellar mass growth in M31-mass and MW-mass progenitors can be
accounted for by the measured SFR evolution from the UV and IR
observations.  Therefore, there is little room for \textit{ex situ}
mass (i.e., mass in stars accreted directly in small satellites) to
contribute to the stellar-mass evolution for present-day $M^\ast$
galaxies at least for $z \lsim 2-3$.     Similar results are also
derived independently by \citet{vandokkum13} \ed{and \citet{patel13b}}, and
are similar to the findings by \citet{moster13} and \citet{behr13a}
based on independent abundance matching methods for dark-matter halos
hosting present-day $M^\ast$ galaxies.  
 
\subsubsection{Quiescent Transition Phase}

\ifsubmode
\begin{figure}
\epsscale{0.8}
\else
\begin{figure}
\epsscale{1.1}
\fi
\plotone{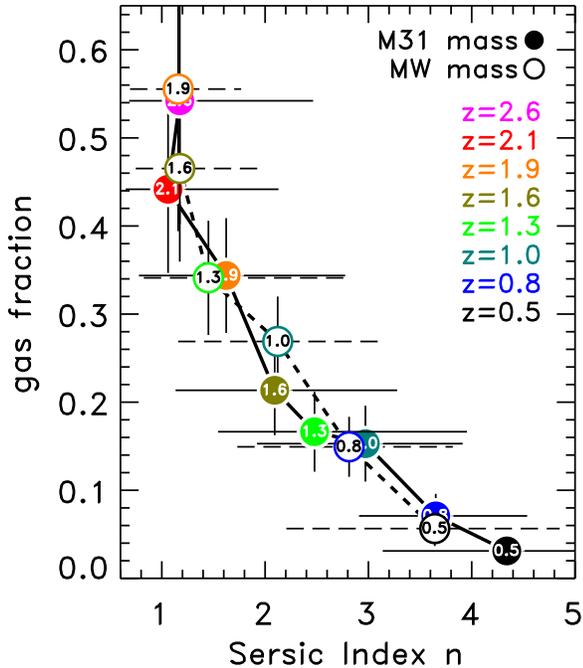}
\caption{Correlation between the \sersic\ index and the implied gas
fraction.     The measurements for the M31 progenitors are shown as
filled circles and solid lines.  The measurements for the MW
progenitors are shown as open circles and dashed lines. Redshifts are
as labeled.  There is no difference between the curves, even though they
are offset in redshift (see figure~\ref{fig:gasevol}), and therefore
the decline in the gas fraction is correlated with the increase in
galaxy bulges.    }\label{fig:gasSersic} \epsscale{1} \ifsubmode
\end{figure}
\else
\end{figure}
\fi
 
During the luminous IR galaxy phase, some of the progenitors of
MW-mass and M31-mass galaxies remain in stages of dust-obscured
star-formation, but the quiescent fraction steadily grows with
decreasing redshift (fig.~\ref{fig:qfrac}).   The increase in the
fraction of quiescent galaxies is simultaneous with an increase with
the morphological \sersic\ index, where the quiescent fraction reaches
$\approx$50\% when the  \sersic\ index reaches $\nsersic\simeq 2$.  As
the quiescent fraction and \sersic\ index increase, the implied gas
fraction decreases.  The anticorrelation between the \sersic\ index
and star-formation activity is consistent with other findings between
star-formation and increasing stellar-mass surface density and bulge
growth \citep[\eg,][]{kauf03b,franx08,bell12,papo12,bass13,patel13a}.
Figure~\ref{fig:gasSersic} shows the evolution of the \sersic--index
against the implied gas fraction for the $M^\ast$ progenitors.
The evolution is nearly indistinguishable for the MW- and
M31-mass progenitors. Therefore, the redshift-dependent differences
in the evolution of $M^\ast$ galaxies divides out when considering the
\sersic\ index--gas fraction plane.   This is evidence that
the formation of bulges coincides with the decline in the gas
fraction. 

Lastly, it seems unlikely that major mergers dominate the assembly of
$M^\ast$ galaxies as they move through the quiescent transition phase.
If major mergers are frequent, then we could expect larger scatter in
colors of the M31- and MW-mass progenitor galaxies (see
fig.~\ref{fig:colormassz}), unless the timescales for these are so short that the
median color evolution remains smooth.  
% 
%There are also few indications of mergers in
%the galaxy morphologies (see figs.~\ref{fig:imgAndro} and
%\ref{fig:imgMW}), but 
%
 Several studies found that major mergers seem
insufficient to produce the observed abundance of spheroidal galaxies
\citep[including, for example,][]{jogee08,bundy09,stew09b}.
Additional evidence disfavoring major mergers comes from the facts
that the mass growth from star-formation accounts for the stellar mass
growth from abundance matching \citep[fig.~\ref{fig:bc},
and][]{moster13,behr13a}), and approximately accounts for all the mass
evolution from constant number density (see fig.~\ref{fig:ndense}).   
 
It remains an open question why, even at fixed stellar mass, some
fraction of the $M^\ast$ progenitors at fixed redshift are star-forming
and some are quiescent galaxies. This is similar to the puzzle
observed in \citet{kawi14}, who found differences in the number of
satellites between star-forming and quiescent central galaxies matched in
stellar mass and redshift $1 < z < 3$.  This implies that at fixed
mass quiescent galaxies have larger-mass halos, and Kawinwanichakij et
al.\ find that a single halo--mass quenching threshold is unable to
reproduce the galaxy distributions.  One conclusion is that while halo mass
is the most important driver of galaxy growth and assembly, other
factors also play a role (such as environment, local galaxy density,
mergers) in determining if a galaxy at fixed stellar mass is quiescent or
remains star-forming \citep[see also, \eg,][]{watson14}.

\subsection{Bulge Formation and Morphological Quenching}

At $z\lsim 1$, during the quiescent-transition phase,  the $M^\ast$
galaxies appear to develop bulge/spheroidal components as evidenced by
their visual morphology (figures~\ref{fig:imgAndro} and
\ref{fig:imgMW}), and their higher \sersic-indices at these redshifts
(figure~\ref{fig:n}).  While this process may start as early as $z\sim
2$ for the M31-mass progenitors, this may depend partially on
environment.    In contrast, there is a marked absence of
bulge-dominated ($\nsersic > 2$) progenitors at high redshift, as was
noted in other studies of MW-sized progenitors
\ed{\citep[\eg][]{patel13b,vandokkum13}}.
%
%  It appears that the
%bulges and spheroidal components of present-day $M^\ast$ galaxies do
%not form prior to the formation of disks (or we would see more
%evidence of high-\sersic-index objects at $ z > 2$).   

The emergence of bulges and spheroids occurs during or after the periods of most
active star-formation.  Typical $M^\ast$-progenitor galaxies have
$\nsersic \simeq 1$ at early times (highest redshifts).  At lower
redshifts, we observe on average monotonic increase in the \sersic\
index with decreasing redshift starting at $z\sim 2$ for the M31
progenitors and $z\sim 1.5$ for the MW progenitors that evolves as
$\nsersic \propto (1+z)^{-2}$.  
%
%This decline in the SFR implies the gas fraction is also declining, and this decline
%correlates strongly with the increase in the \sersic\ index 
%which is illustrated in figure~\ref{fig:gasevol}.  The
%process of bulge/spheroidal formation is gradual: figure~\ref{fig:n}
%

\citet{bruce12} show that $\nsersic=2$ (3)
corresponds approximately to a bulge stellar--mass fraction
(bulge-to-total, B/T) of 50\% (75\%).   Therefore, on average M31- and
MW-mass galaxies develop a significant bulge ($\nsersic=2$, with
presumably B/T$=50$\%) by $z\simeq 1.7$ and 1.1, respectively; and
they develop a dominant bulge ($\nsersic=3$, with B/T=75\%) by
$z\simeq 1.0$ and 0.7, respectively.  However, there is appreciable
scatter about the average, and this is likely related to the fact that
at any fixed redshift the progenitor population is a mix of
star-forming and quiescent galaxies (where the latter presumably have
higher $\nsersic$).  
%These differences between quiescent and
%star-forming MW-mass progenitors will be discussed in more
%detail in K.~Sharon et al.\ (2014, in preparation).  

%To summarize, if we associate the formation and emergence of galaxy
%bulges (with B/T $\sim$ 50--75)\% with \sersic\ index evolution from
%$\nsersic=2-3$, then bulges emerge between $z=1.6$ and 1 for the
%M31-mass progenitors and $z=$1.3 and 0.7 for the MW-mass progenitors.
%There is a clearly a strong dependence in bulge formation and the
%stellar mass. In both cases bulge emergence corresponds to the decline in the
%overall galaxy SFR.  
%
%

%
%Relating increased \sersic\ indices to an increase in the
%stellar-mass surface density (for example, at fixed mass the galaxy
%with higher \sersic\ index will have higher stellar mass surface
%density), it then seems that most of the evolutionary differences in
%the $M^\ast$ progenitors can be attributed to their difference in
%stellar-mass surface density rather than purely their stellar mass. 

The formation of spheroids/bulges in the $M^\ast$ progenitors samples
is tied to the transition from star-forming to quiescent phases of the
galaxies' evolution.   Recent theoretical models connect the formation
of stellar spheroids to the dynamics and quenching of star-formation
in gas-rich, turbulent disks.  In these ``morphological'' quenching
(MQ) models the formation of stellar spheroids and stellar-dominated disks
stabilize the gas in the galactic disks, suppressing the
star-formation efficiency
\citep[\eg,][]{bour07a,elme08,dekel09b,martig09,martig10,ceve10,genel12,sales12,zava12},
and this suppression may be enhanced through an increase in the
velocity dispersion of the gas arising from stellar and AGN feedback
\citep[also called ``$Q$'' quenching, where the increase in velocity
dispersion lowers the Toomre $Q$-parameter, see][]{dekel14}.  

Other observational evidence supports the notion that the MQ and
$Q$-quenching processes occur and that quiescent galaxies contain cold
reservoirs.  \citet{kenn89} showed that locally, bulge dominated
galaxies form stars with lower efficiency than disk-dominated ones.
\citet{tuml13} detected large column densities of cold gas (with low
star-formation efficiency) in sightlines through the halos of
quiescent galaxies.  And, \citet{genzel14} provided evidence that this
type of quenching occurs in the outer disks of massive, star-forming
disk galaxies at high redshift with centrally concentrated mass
distributions.  This is also consistent with the observations by
\citet{vandokkum13} \ed{and \citet{patel13b}} who find that galaxies
in their sample of MW-mass progenitors form stars in disks at all
radii, at least for $z > 0.6$, broadly consistent with expectations
from MQ and $Q$-quenching theories.  

The MQ and $Q$-quenching models are consistent with the
anticorrelation between the specific SFR and \sersic\ index for the
$M^\ast$-galaxy progenitors.  Some MQ and $Q$-quenching models predict
that the galaxies should contain cold gas with low star-formation
efficiency \citep[\eg,][]{dekel09b}.  This prediction is in line with
our results, and plausibly accounts for the discrepancy between the
(higher) observed implied gas masses for the $M^\ast$-galaxy
progenitors and the gas masses measured for local galaxies directly
by, e.g., \citet{sain11}.   To test these predictions further requires
direct measurements of the gas masses of the $M^\ast$ progenitors at
high redshift.  How the gas fractions correlate with the \sersic\
indices will lead to an understanding of bulge formation and how it
affects galaxy formation.  This should be testable in part for our
sample through future observations from ALMA.

\subsection{The Dependence on Stellar Mass}

Based on all the observables in this paper, the progenitors of
M31-mass and MW-mass galaxies go through all the same stages of
evolution, but the higher mass M31-mass progenitors advance through
the stages at earlier times (i.e., at higher redshifts) than the
lower-mass MW-mass progenitors.   Furthermore, the M31-mass
progenitors have higher stellar masses at the time they advance to the
later evolutionary stages compared to the MW-mass progenitors.  It
is somewhat remarkable that the M31- and MW-mass progenitors exhibit
these evolutionary differences because their present-day stellar
masses are separated by only a factor of two.  The most obvious
explanation for why the M31-mass progenitors begin evolutionary stages
sooner is that they have larger halo mass at earlier times
\citep{moster13,behr13a}.   
%
%However, it seems that halo mass may not
%the whole story as at fixed \textit{stellar mass} the MW-mass
%progenitors have evolved to later stages compared to the M31-mass
%progenitors. 
%
 If the evolution is driven by the evolution of the mass
of the dark-matter halo only, then the stellar-mass/halo-mass ratio
evolves strongly with halo mass and redshift. Alternatively, some
other process must also be involved. 
%
%While there
%is some evidence to support this interpretation, mass is not the whole
%story, primarily because as the galaxies grow, the Universe also
%expands, which lowers the gas-accretion rate onto halos. 

At the highest redshifts to which we are able to observe the $M^\ast$
galaxy progenitors ($z\sim 2.5-3)$, the SFR and stellar mass grow in
tandem.  This keeps the specific SFR roughly constant (see
figure~\ref{fig:lirevol}).   At this point the halos in which the
$M^\ast$ progenitors reside accrete gas from the intergalactic medium, and both
empirical constraints \citep[\eg,][]{papo11} and theoretical
predictions favor models where the gas accretion is comparable to the
SFR \citep[\eg,][]{neis06,agertz09,bouche10,ceve10,dekel13,dekel14b}.  
The $M^\ast$ galaxies remain in these ``steady-state''
\citep[\eg,][]{dekel13} phases until $z\simeq 2.5$ (1.5) for the M31-
(MW-)mass progenitors, where the SFR
and sSFR then decline at lower redshifts.   The decline in the sSFR is
similar in form for the two sets of progenitors.  However, the
decline begins for the M31-mass progenitors when their stellar masses
are between $\log M_\ast/M_\odot = 10.2$ and 10.4, a factor of 2-3
\textit{higher} than stellar mass of the MW-mass progenitors.  This is
exactly the same effect as seen in the difference in the evolution of
the rest-frame colors and mass for the progenitors of the M31-mass and
MW-mass galaxies observed in figure~\ref{fig:colormassz}.  

The process that begins the suppression of the SFR in $M^\ast$
progenitors can not be entirely driven by a simple unevolving
halo-mass threshold unless the stellar-mass/halo-mass relation evolves
strongly with halo mass and redshift.   One possibility is that as the galaxies grow the
Universe also expands, such that by the time the smaller MW-mass
galaxies reach a target stellar mass, the cosmic baryon (and
dark-matter) density is lower, which lowers the gas-accretion rate
onto halos \citep[see \eg,][]{behr13b}.   This is conceptually similar
to models that find the quenching halo-mass threshold is higher at
higher redshift \citep{dekel06a}, and this may drive
evolution in the stellar-mass/halo-mass relation
\citep{moster13,behr13c}. 

Therefore, models for the evolution of the star-formation efficiency
and quenching in galaxies require both a redshift and mass dependence.
The mass dependent star-formation efficiency is easy to understand in
terms of feedback processes, which are stronger when halos are lower
mass with shallower gravitational potential wells
\citep[\eg,][]{luyu14a}.   Another solution may be that gas
in the halos of galaxies is preheated by early starbursts or other
processes, which delays baryonic accretion onto the galaxies
\citep{luyu14b}.  This could  allow for redshift-dependent effects in
the star-formation efficiency at fixed halo mass.   The
redshift-dependence effects would also account for why the lower-mass
MW-mass galaxies never attain the same peak SFRs as the higher mass
M31-mass galaxies.

\section{Summary}

We track the evolution of progenitors of present-day $M^\ast$
galaxies, selected from abundance-matching methods of $z=3$ to 0.5.
The abundance-matching methods account effects of galaxy formation and
mergers.   We track the evolution of $M^\ast $ galaxies at two
present-day values of stellar mass, including M31-mass progenitors
with present-day stellar masses of $10^{11}$~\msol\ and MW-mass
progenitors with present-day stellar mass of $5\times 10^{10}$~\msol.
This allows us to study the mass-dependent evolution for a full range
of stellar masses that encompass present-day $M^\ast$ galaxies,
including galaxies like the MW and M31 proper.  Furthermore, we are
able to study the evolution of galaxies separated by a factor of two
in stellar mass.

The data paint a consistent picture of present-day $M^\ast$ galaxy
evolution, and this is based on three independent datasets:  the
rest-frame $U-V$, $V-J$ color evolution derived from the full $0.3 -
8.0$~\micron\ photometric datasets,  the evolution of sizes and
\sersic\ indices from \hst/WFC3 imaging, and evolution in the far-IR
luminosities as measured from the \spitzer\ and \herschel\ data.
There is appreciable scatter in the observed quantities for the
progenitors at any redshift \citep[and partly this is because the
progenitors of a present-day $M^\ast$ galaxy have a range of mass at
higher redshift, see][]{behr13c}.    Therefore the
evolution processes and timescales clearly vary from galaxy to
galaxy, and may not apply to individual $M^\ast$ galaxies.  However,
for the population, the average evolution of the ``typical''
progenitor is still very enlightening for the formation of the
present-day $M^\ast$ galaxy population.  

All the progenitors of the present-day $M^\ast$ galaxies go through
the same evolutionary stages:

\begin{enumerate}
\item  At the earliest epochs, the progenitors are blue and
star-forming, with relatively unattenuated galaxies, with properties
similar to LBGs and LAEs.  This is therefore the ``LBG phase''.    The
morphologies have disk-like exponential surface-brightness profiles
(\sersic\ index $\nsersic=1$) that grow as $H(z)^{-1}$, as expected
for the smooth growth of the dark-matter halos (at least under the
assumption of a constant halo spin parameter).    For the M31-mass
progenitors this phase extends to $z \gsim 2.5$, and for the MW-mass
progenitors this phase occurs at $z\gsim 2$.  

\item At later times, the ``typical'' progenitor becomes an IR-luminous
star-forming galaxies with higher dust obscuration.  This is observed
in the evolution of the median rest-frame colors and the median IR
luminosity.    For M31-mass progenitors this phase begins around
$z=2.5$ and continues to $z\sim 1.5$, with a peak SFR of $\Psi =
50$~\msol\ yr$^{-1}$ from $z\sim 2.5$ to $z\sim 1.8$.  For MW-mass
progenitors the IR luminous phase begins later, from $z\sim 1.8$ to
$z\sim 1.0$, with a peak SFR of $\Psi = 30$~\msol\ yr$^{-1}$ around
$z\sim 1.5$.  This is the ``Luminous IR galaxy'' phase of
$M^\ast$-galaxy evolution. 

\item During the ``Luminous IR galaxy'' phase the
\sersic\ index increases from $\nsersic\simeq 1$ at a rate of  roughly
$\nsersic\propto (1+z)^{-2}$ starting at $z\sim 2$ for the M31-mass
galaxies and $z\sim 1.5$ for the MW-mass galaxies.   \ed{The $M^\ast$
galaxies appear to populate the well-known ``Hubble Sequence'' by
$z\sim 1$.}  The fraction of
quiescent galaxies also rises with decreasing redshift in tandem with
the \sersic\ index evolution (and therefore bulge growth) such that
when $\nsersic\simeq 2$ the quiescent fraction is approximately 50\%.
This is accompanied with a decline in the specific SFR.  

\item At the latest times (lowest redshifts), the SFR for
both the M31-mass and MW-mass progenitors show  a steady decline, most
pronounced for $z < 1$.         During this period the \sersic\
indices continue to increase with decreasing redshift and the
quiescent fraction increases such that the typical M31-mass progenitor
has $\nsersic=3.6$ with a quiescent fraction of 70\% at $z=0.8$.  The
typical MW-mass progenitor evolution is delayed and it reaches these
values at $z\sim 0.5$.  This stage is the ``quiescent transition''
phase.  The majority of present-day $M^\ast$ galaxies
should be quiescent, and this seems consistent with observations
\citep[although it may be that the MW and M31 proper are outliers,
see][]{mutch11}. 

\end{enumerate}

While the M31-mass and MW-mass progenitors experience the same
evolutionary phases, the M31-mass galaxies experience them sooner
(higher redshift) with higher relative stellar mass compared to the
MW-mass progenitors.   This means the threshold mass for any quenching
process depends on redshift (and therefore the stellar-mass/halo-mass
relation is evolving) and may be related to physical processes
motivated by models, including lower gas accretion rate, an evolving
magnitude of galaxy feedback, preheating of baryonic gas in
galaxy halos, and lower cosmic density of baryons as the Universe
evolves. 

Our observations show that the formation of bulges in $M^\ast$
galaxies (as measured from the \sersic\ indices) is simultaneous with a
decline in the inferred  cold gas mass (implied by inverting the
Kennicutt--Schmidt law), and suggests these processes are related.
Because the evolution in these quantities is smooth for the
``typical'' $M^\ast$-galaxy progenitor, this favors slower acting
processes that drive the transition from gas-rich (star-forming) disk
galaxy to quiescent, bulge-dominated galaxy, rather than dramatic,
stochastic processes, like major mergers, as a dominant driver for
this evolution, unless the timescales for the latter are so short that the
median evolution remains smooth.  Our observations are consistent with
ideas in morphological-quenching and $Q$-quenching models.  These
predict generically that the growth of stellar bulges and disks
stabilize cold gas against fragmentation, perhaps combined with an
increase in gas turbulence from stellar/AGN feedback, which lowers the
Toomre $Q$ parameter.  

One test for understanding the evolution of $M^\ast$ progenitors will
come from measurements of the cold-gas mass (and the
gas-mass/stellar-mass ratio) from observations with facilities such as
ALMA.  If the gas fractions are as low as implied here from the SFR
and sizes of the galaxies, then the processes driving the transition
of a star-forming galaxy to a quiescent, bulge-dominated galaxy must
remove the gas from the systems.  In contrast, if processes stabilize
the gas against instabilities, then the galaxies will have higher
measured cold-gas fractions with low star-formation efficiencies,
which will favor morphological-quenching and $Q$-quenching models.
Correlating these cold-gas measurements with galaxy morphological
properties (such as the bulge/total ratio) will test the details of
these theories. 

%The formation of galaxy spheroids (bulges) with \sersic\ indices $n >
%2$, then bulges emerge for the M31-mass and MW-mass galaxies at
%$z\sim 1$ and 0.7, respectively.    The formation of the bulges
%appears to correspond to the phase of high IR lumosities, and
%therefore stellar mass growth and bulge formation appear to occur
%concurrently, where a significant bulge emerges at the end of this
%period.  
%
%Lastly, we note that major mergers probably play only a small role in
%the formation of present-day $M^\ast$ galaxies from $z\sim 3$ to the
%present.   Therefore,
%major mergers play a small role in the mass assembly of $M^\ast$
%galaxies. 
%

\acknowledgments

\ed{We thank our other colleagues in CANDELS and ZFOURGE for their
work and collaboration.  We thank B.\ Moster for valuable input, and
we thank the referee for a thorough report that improved the quality
and clarity of this paper.}  This work is supported by the National
Science Foundation through grant AST-1009707.  \ed{IL acknowledges
support from ERC HIGHZ \#227749 and NL-NWO Spinoza.}  This work is
based on observations taken by the CANDELS Multi-Cycle Treasury
Program with the NASA/ESA HST, which is operated by the Association of
Universities for Research in Astronomy, Inc., under NASA contract
NAS5-26555.  This work is supported in part by HST program number
GO-12060.  Support for Program number GO-12060 was provided by NASA
through a grant from the Space Telescope Science Institute, which is
operated by the Association of Universities for Research in Astronomy,
Incorporated, under NASA contract NAS5-26555.  This work is based on
observations made with the \textit{Spitzer Space Telescope}, which is
operated by the Jet Propulsion Laboratory, California Institute of
Technology.  This work is based on observations made with the
\textit{Herschel} Space Observatory.  \textit{Herschel} is an ESA
space observatory with science instruments provided by European-led
Principal Investigator consortia and with important participation from
NASA.  This paper includes data gathered with the 6.5 meter Magellan
Telescopes located at Las Campanas Observatory, Chile.  Australian
access to the Magellan Telescopes was supported through the National
Collaborative Research Infrastructure Strategy of the Australian
Federal Government.  We acknowledge generous support from the Texas
A\&M University and the George P.\ and Cynthia Woods Institute for
Fundamental Physics and Astronomy.   

\ed{This paper is dedicated to the memory of
Alma J.\ Broy and James W.\ Broy of Baltimore, Maryland.}

\medskip

\bibliography{apj-jour,alpharefs}{}
\bibliographystyle{apj}

\end{document}